\documentclass[11pt,a4paper,notcite,notref]{article}

\usepackage{amssymb,amsmath}
\usepackage{tipa}
\usepackage{epic}
\usepackage{amsfonts}
\usepackage{accents}
\usepackage{texdraw}
\usepackage{color}
\usepackage{eucal}
\usepackage{cite}
\usepackage[all]{xy}
\usepackage{bm}
\usepackage{graphics,amsmath,amssymb,amsthm,accents}
\newtheorem{Proposition}{Proposition}
\numberwithin{equation}{section}

\def \tyb#1{\hbox{\tiny{[{\it{#1}}]}}}
\def \ty#1{\hbox{\tiny{{\it{#1}}}}}

\def \dh#1{\underaccent{\hat}{#1}}
\def \dt#1{\underaccent{\tilde}{#1}}
\def \db#1{\underaccent{\bar}{#1}}

\DeclareMathAccent{\wtilde}{\mathord}{largesymbols}{"65}
\DeclareMathAccent{\what}{\mathord}{largesymbols}{"62}

\def\m@th{\mathsurround=0pt}
\mathchardef\bracell="0365
\def\upbrall{$\m@th\bracell$}
\def\undertilde#1{\mathop{\vtop{\ialign{##\crcr
    $\hfil\displaystyle{#1}\hfil$\crcr
     \noalign
     {\kern1.5pt\nointerlineskip}
     \upbrall\crcr\noalign{\kern1pt
   }}}}\limits}

\def\underhat#1{\mathop{\vtop{\ialign{##\crcr
    $\hfil\displaystyle{#1}\hfil$\crcr
     \noalign
     {\kern1.5pt\nointerlineskip}
     \upbrall\crcr\noalign{\kern1pt
   }}}}\limits}

\newcommand{\wb}[1]{\overline{#1}}

\newcommand{\wh}{\widehat}
\newcommand{\wt}{\widetilde}
\newcommand{\ut}{\undertilde}

\newcommand{\ts}{\,^t \hskip -2pt {\boldsymbol{s}}}
\newcommand{\tu}{\,^t \hskip -2pt {\boldsymbol{u}}}

\newcommand{\Ga}{\boldsymbol{\Gamma}}
\newcommand{\Lb}{\boldsymbol{\Lambda}}
\newcommand{\sg}{\sigma}
\newcommand{\ka}{\kappa}
\newcommand{\vi}{\varpi}

\newcommand{\var}{\varphi}

\newcommand{\ci}{\chi}
\newcommand{\vs}{\varsigma}
\newcommand{\ro}{\rho}
\newcommand{\vr}{\varrho}

\newcommand{\nn}{\nonumber}

\newcommand{\bA}{\boldsymbol{A}}
\newcommand{\bB}{\boldsymbol{B}}
\newcommand{\bC}{\boldsymbol{C}}

\newcommand{\bF}{\boldsymbol{F}}
\newcommand{\bG}{\boldsymbol{G}}
\newcommand{\bH}{\boldsymbol{H}}
\newcommand{\bI}{\boldsymbol{I}}

\newcommand{\bK}{\boldsymbol{K}}
\newcommand{\bL}{\boldsymbol{L}}
\newcommand{\bM}{\boldsymbol{M}}

\newcommand{\bT}{\boldsymbol{T}}

\newcommand{\br}{{\boldsymbol{r}}}
\newcommand{\bu}{\boldsymbol{u}}
\newcommand{\st}{\hbox{\tiny\it{T}}}

\begin{document}

\title{Generalized Cauchy matrix approach for non-autonomous discrete Kadomtsev-Petviashvili system}

\author{Songlin Zhao\footnote{Corresponding author.
E-mail:~songlinzhao1984@gmail.com}~,~~Wei Feng,~~Shoufeng Shen,~~Jun Zhang
\vspace{1mm}\\
{\small \it Department of Applied Mathematics, Zhejiang University of Technology, Hangzhou 310023, China}}

\date{}
\maketitle
\begin{abstract}
In this paper, we investigate the non-autonomous discrete Kadomtsev-Petviashvili (KP) system
in terms of generalized Cauchy matrix approach. These equations include non-autonomous bilinear lattice
KP equation, non-autonomous lattice potential KP equation,
non-autonomous lattice potential modified KP equation,
non-autonomous asymmetric lattice potential modified KP equation,
non-autonomous lattice Schwarzian KP equation and non-autonomous
lattice KP-type Nijhoff-Quispel-Capel equation. By introducing point transformations,
all the equations are described as simplified forms, where
the lattice parameters are absorbed. Several kinds of solutions more than multi-soliton solutions to these equations
are derived by solving determining equation set. Lax representations for these equations are also discussed.

\vspace{0.5cm} \noindent {\bf Keywords}\quad Non-autonomous discrete KP system,
Generalized Cauchy matrix approach, Exact solutions, Lax representations

\vspace{0.5cm} \noindent {\bf PACS numbers:} \quad  02.30.Ik,
02.30.Ks, 05.45.Yv

\end{abstract}

\section{Introduction}

As one of the most important models in nonlinear evolution equation, KP equation has attracted lots of
attention. The first discrete KP equation, which in the form of bilinear structure, was proposed by Hirota in Ref. \cite{H-BLKP},
where the corresponding soliton solutions, Lax pair and B\"{a}cklund transformation were also studied.
Subsequently, operator approach \cite{Miwa}, Lie-algebraic approach \cite{D-1982}, direct linearization (DL)
method \cite{N-KP-DL}, Cauchy matrix approach \cite{N-KP} and its generalization \cite{WZ} were also used to construct the
discrete KP system. In Ref. \cite{ABS-2012}, Adler, Bobenko and Suris (ABS) utilized consistency approach to
classify discrete integrable three-dimensional
equations of the octahedron type, in which discrete KP equation and its Schwarzian version were
included and discrete KP equation was shown 4D-consistency on a 4D cube.
The discrete KP system is comprised of several equations, which are bilinear lattice KP (blKP) equation,
lattice potential KP (lpKP) equation, lattice potential modified KP (lpmKP) equation,
asymmetric lpmKP equation, lattice Schwarzian KP (lSKP) equation
and lattice KP-type Nijhoff-Quispel-Capel (lKP-NQC) equation. Here we also call these equations: lattice KP-type equations.
All the above equations in discrete KP family can be described as
\begin{align}
Q(\wt{f},~\wh{f},~\wb{f},~\wh{\wt{f}},~\wb{\wt{f}},~\wb{\wh{f}};~p,q,r)=0,
\label{Q}
\end{align}
where $f:=f_{n,m,h}=f(n,m,h)$ denotes the dependent variable of the lattice
points labeled by $(n,m,h)\in \mathbb{Z}^3$; $p$, $q$ and $r$ are continuous lattice parameters associated
with the grid size in the directions of the lattice given by the
independent variables $n$, $m$ and $h$. In equation \eqref{Q}, we have employed the notations
\[\wt{f}=f_{n+1,m,h},~\wh{f}=f_{n,m+1,h},~\wb{f}=f_{n,m,h+1},\]
in terms of which we also have
\[ \wh{\wt{f}}=f_{n+1,m+1,h}, ~ \wt{\wb{f}}=f_{n+1,m,h+1}, ~ \wh{\wb{f}}=f_{n,m+1,h+1}. \]
In addition, we define the backward direction of shifts $\wt{\phantom{a}}$, $\wh{\phantom{a}}$,
and $\wb{\phantom{a}}$ as
\[\ut{f}=f_{n-1,m,h},~\dh{f}=f_{n,m-1,h},~\db{f}=f_{n,m,h-1}.\]

When parameters $p,q$ and $r$ in \eqref{Q} are constants, equation \eqref{Q} is called autonomous equation,
while when $p,q$ and $r$ are defined by $p=p(n)=p_n,q=q(m)=q_m$ and $r=r(h)=r_h$, equation \eqref{Q}
corresponds to the non-autonomous equation. In recent two decades, the study of the non-autonomous discrete integrable systems has been
a hot topic and some significant progress has been made.
In Ref. \cite{SC}, Sahadevan and Capel utilized Lax pair technique and singularity confinement criteria to study
the complete integrability of the non-autonomous discrete modified Korteweg-de Vries and sine-Gordon mappings.
Besides, with the help of the Lax pair approach, the ultra-local singularity
confinement criterion and direct construction of conservation laws, Sahadevan {\it et al.}
also discussed the integrability conditions for non-autonomous quad-graph equations \cite{SRH}.
Grammaticos and Ramani \cite{GR} investigated the non-autonomous ABS lattice by
singularity confinement and algebraic entropy approach.
With regard to the solutions, Willox and his collaborators constructed solutions for non-autonomous
blKP equation by Darboux transformation (DT) and binary DT \cite{WTS,WTS-1}. Hay gave the
Casorati determinant solutions to the non-autonomous cross-ratio equation \cite{H-cr}.
Kajiwara {\it et al.} derived soliton solutions for many non-autonomous discrete equations in terms of
bilinear formalism \cite{KM,KO-1,KO-2}, including discrete KP hierarchy.
Recently, Zhang and his collaborators investigated exact solutions to several non-autonomous discrete equations by
means of bilinear method, including non-autonomous H1, H2, H3 and Q1 in non-autonomous ABS lattice \cite{SZZ} and
non-autonomous lattice Boussinesq equation \cite{NZ}.

Motivated by the generalized Cauchy matrix approach for the autonomous lattice KP-type equations \cite{WZ},
in present paper we will make use of this method to study the
non-autonomous lattice KP-type equations, including non-autonomous blKP equation,
non-autonomous lpKP equation, non-autonomous lpmKP equation, non-autonomous asymmetric lpmKP equation,
non-autonomous lSKP equation and non-autonomous lKP-NQC equation. A determining equation set
(DES) will be firstly introduced, based on which evolution of matrix $\bM$ will be given. Furthermore,
by defining scalar function $S^{(i,j)}(a,b)$ and considering its evolution, recurrence relations
of $S^{(i,j)}(a,b)$ will be constructed, by which the non-autonomous
lattice KP-type equations will be constructed as closed forms
with special choices of $i,j$ and $a,b$. The corresponding exact solutions will be derived
by solving the canonical form of the DES.
As a concrete spin-off of the derivations presented here, we also derive the Lax
representations for the non-autonomous lattice KP-type equations.

The paper is organized as follows. In Section 2, we set up DES and introduce scalar function $S^{(i,j)}(a,b)$
together with vector functions $\bu^{(i)}(a)$ and $\tu^{(j)}(b)$. By considering the evolution of $S^{(i,j)}(a,b)$,
non-autonomous lattice KP-type equations are constructed. The deformation of the non-autonomous lattice KP-type equations
are also derived by imposing point transformations. In Section 3, we discuss exact solutions for the DES.
Sec. 4 is devoted to the construction of Lax representation. Conclusion will be given in the last section.
In addition, we have an Appendix consisting of 2 sections
as a compensation of the paper.

\section{Generalized Cauchy matrix approach} \label{GCMA}

In this section, we will establish the generalized Cauchy matrix approach for the
non-autonomous lattice KP-type equations. Firstly, we will introduce the DES. Secondly,
by proposing scalar function $S^{(i,j)}(a,b)$ and considering its evolution relation,
several non-autonomous lattice KP equations will be constructed as
closed-forms. $\tau$-function will also be considered to build the non-autonomous blKP equation.

In generalized Cauchy matrix approach, the following proposition \cite{S-1884} is always needed.
\begin{Proposition} \label{solv}
Let us denote the eigenvalue sets of matrices $\bA$ and $\bB$ by $\mathcal{E}(\bA)$ and $\mathcal{E}(\bB)$, respectively.
For the known matrices $\bA, \bB$ and $\bC$, equation
\begin{equation}
\bA\bM-\bM\bB=\bC
\label{SE}
\end{equation}
has a unique solution $\bM$ if and only if $\mathcal{E}(\bA)\bigcap \mathcal{E}(\bB)=\varnothing$.
\end{Proposition}
The equation \eqref{SE} is one of the famous matrix equations, named as Sylvester equation. This equation
plays a central role in many areas of applied mathematics. With some more conditions on
$\mathcal{E}(\bA)$ and $\mathcal{E}(\bB)$, solution $\bM$ of \eqref{SE} can be expressed
via series or integration \cite{BR-BLMS-1997}.

\subsection{DES}

Let us start from the following DES
\begin{subequations}
\begin{eqnarray}
&& \bK\bM+\bM\bL=\br\ts, \label{ce-1} \\
&& p_n\wt{\br}=(p_n\bI_N+\bK)\br,~~q_m \wh{\br}=(q_m\bI_N+\bK)\br,~~r_h \overline{\br}=(r_h\bI_N+\bK)\br, \label{ce-2-a}\\
&& \wt{\ts}(p_n\bI_{N'}-\bL)=p_n\ts,~~\wh{\ts}(q_m\bI_{N'}-\bL)=q_m\ts ,~~ \wb{\ts}(r_h\bI_{N'}-\bL)=r_h\ts, \label{ce-2-b}
\end{eqnarray}
\label{ce}
\end{subequations}
where $\bI_N$ and $\bI_{N'}$ are, respectively,
the $N$th-order and $N'$th-order unit matrices;
$\bM\in \mathbb{C}_{N\times N'}$, $\br\in \mathbb{C}_{N\times 1}$ and
$\ts\in \mathbb{C}_{1\times N'}$ are undetermined functions depending on independent variables $n,~m$
and $h$, while $\bK$ and $\bL$ are non-trivial constant matrices. It is easy to recognize that \eqref{ce-1}
is the Sylvester equation corresponding to $\bC$ being rank 1 in \eqref{SE}. To guarantee the Sylvester equation
\eqref{ce-1} can be solved, we assume $\mathcal{E}(\bK)\bigcap \mathcal{E}(-\bL)=\varnothing$.
Besides, we also suppose $s\bI_N\pm \bK$ and $s\bI_{N'}\pm \bL$ are invertible for $s=0,p_n,q_m,r_h,a,b$.

From \eqref{ce}, the evolution of $\bM$ can be obtained. In fact, subtracting \eqref{ce-1} from \eqref{ce-1}$\wt{\phantom{a}}$,
we get
\begin{eqnarray}
\bK(\wt{\bM}-\bM)+(\wt{\bM}-\bM)\bL=\wt{\br}\wt{\ts}-\br\ts.
\label{M-dyna-cal-1}
\end{eqnarray}
Taking the first relations of \eqref{ce-2-a} and \eqref{ce-2-b} into \eqref{M-dyna-cal-1} yields
\begin{eqnarray}
\bK(\wt{\bM}-\bM)+(\wt{\bM}-\bM)\bL=\frac{1}{p_n}\bK\br\wt{\ts}+\frac{1}{p_n}\br\wt{\ts}\bL, \label{M-dyna-cal-2}
\end{eqnarray}
which gives rise to the evolution of $\bM$ in $n$-direction
\begin{eqnarray}
\wt{\bM}=\bM+\frac{1}{p_n}\br\wt{\ts} \label{M-dyna-a}
\end{eqnarray}
in the light of Proposition \ref{solv}.
Replacing $p_n$ by $q_m$, and $\wt{\phantom{a}}$-shift by
$\wh{\phantom{a}}$-shift or $p_n$ by $r_h$, and $\wt{\phantom{a}}$-shift
by $\wb{\phantom{a}}$-shift in \eqref{M-dyna-a}, we obtain
\begin{eqnarray}
\wh{\bM}=\bM+\frac{1}{q_m}\br\wh{\ts},~~\wb{\bM}=\bM+\frac{1}{r_h}\br\wb{\ts}. \label{M-dyna-b}
\end{eqnarray}
Relations \eqref{M-dyna-a} and \eqref{M-dyna-b} encode all the information on the dynamics of the matrix $\bM$, w.r.t. the
independent variables $n,m$ and $h$, in addition to \eqref{ce-1} which can be thought as the defining
property of $\bM$.

\subsection{Master function $S^{(i,j)}(a,b)$}

\subsubsection{The definition of $S^{(i,j)}(a,b)$}

Now we introduce scalar function
\begin{eqnarray}
S^{(i,j)}(a,b)&=&\ts(b\bI_{N'}+\bL)^j\bC(\bI_{N}+\bM\bC)^{-1}(a\bI_N+\bK)^i \br \nn \\
&=&\ts(b\bI_{N'}+\bL)^j(\bI_{N'}+\bC\bM)^{-1}\bC(a\bI_N+\bK)^i\br, ~~i,j\in \mathbb{Z},
\label{eq:Sijab}
\end{eqnarray}
where $\bC \in \mathbb{C}_{N'\times N}$ is an arbitrary
constant matrix; parameters $a$ and $b$ are freedom, which can be either constants or functions depending on independent variables
$n,m$ and $h$. Here we still call $S^{(i,j)}(a,b)$ the master function (see Ref. \cite{ZZ}), since it will be
used to generate non-autonomous lattice KP-type equations. In addition to \eqref{eq:Sijab}, we also need the following
auxiliary vector functions
\begin{subequations}
\begin{eqnarray}
&& \bu^{(i)}(a)=(\bI_N+\bM\bC)^{-1} (a\bI_N+\bK)^i \br,~~i\in \mathbb{Z}, \label{uia} \\
&& \tu^{(j)}(b)=\ts(b\bI_{N'}+\bL)^j(\bI_{N'}+\bC\bM)^{-1},~~j\in \mathbb{Z}. \label{tujb}
\end{eqnarray}
\end{subequations}
Then $S^{(i,j)}(a,b)$ can be expressed by $\bu^{(i)}(a)$ and $\tu^{(j)}(b)$ through
\begin{eqnarray}
S^{(i,j)}(a,b)=\ts(b\bI_{N'}+\bL)^j\bC\bu^{(i)}(a)=\tu^{(j)}(b)\bC(a\bI_N+\bK)^i \br. \label{Su-re}
\end{eqnarray}

\subsubsection{Invariance of $S^{(i,j)}(a,b)$} \label{ss-inva}

Supposing that under transformed matrices $\bT_1$ and $\bT_2$, matrices $\bK_1$ and $\bL_1$ are, respectively,
similar to $\bK$ and $\bL$, i.e.
\begin{equation}
\bK_1=\bT_1 \bK \bT_1^{-1},~~\bL_1=\bT_2 \bL \bT_2^{-1}.
\label{KL1-KL}
\end{equation}
Then with transformations
\begin{equation}
\bM_1=\bT_1 \bM \bT_2^{-1},~~\bC_1=\bT_2 \bC \bT_1^{-1},~~\br_1=\bT_1 \br,~~\ts_1=\ts\bT_2^{-1},
\label{Mrs-1}
\end{equation}
DES \eqref{ce} becomes
\begin{subequations}
\begin{align}
& \bK_1\bM_1+\bM_1\bL_1= \br_1\ts_1,\label{Se-inv} \\
& p_n\wt{\br}_1=(p_n\bI_N+\bK_1)\br_1,~~q_m \wh{\br_1}=(q_m\bI_N+\bK_1)\br_1,~~r_h \overline{\br}_1=(r_h\bI_N+\bK_1)\br_1, \label{r-inv} \\
& \wt{\ts}_1(p_n\bI_{N'}-\bL_1)=p_n\ts_1,~~\wh{\ts}_1(q_m\bI_{N'}-\bL_1)=q_m\ts_1,~~ \wb{\ts}_1(r_h\bI_{N'}-\bL_1)=r_h\ts_1, \label{s-inv}
\end{align}
and \eqref{eq:Sijab} yields
\begin{align}
S^{(i,j)}(a,b)&=\ts(b\bI_{N'}+\bL)^j\bC(\bI_{N}+\bM\bC)^{-1}(a\bI_N+\bK)^i\br \nn \\
&=\ts_1(b\bI_{N'}+\bL_1)^j\bC_1(\bI_{N}+\bM_1\bC_1)^{-1}(a\bI_N+\bK_1)^i\br_1.
\end{align}
\end{subequations}
This means that the master function $S^{(i,j)}(a,b)$ is invariant under transformations \eqref{KL1-KL} and \eqref{Mrs-1}.

\subsubsection{Evolution of $S^{(i,j)}(a,b)$}

To proceed, we firstly consider the evolutions of $\bu^{(i)}(a)$ and $\tu^{(j)}(b)$. Function \eqref{uia}
is equivalent to
\begin{eqnarray}
(\bI_N+\bM\bC)\bu^{(i)}(a)=(a\bI_N+\bK)^i \br. \label{uia-dp-cal-1}
\end{eqnarray}
Taking $\widetilde{\phantom{a}}$-shift of \eqref{uia-dp-cal-1} and noting that \eqref{M-dyna-a} and \eqref{ce-2-a}, we have
\[
p_n(\bI_N+\bM\bC)\wt{\bu}^{(i)}(a)=(p_n-\wt{a})(\wt{a}\bI_N+\bK)^i \br+(\wt{a}\bI_N+\bK)^{i+1}\br-\br\wt{\ts}\bC\wt{\bu}^{(i)}(a),
\]
which by relation \eqref{Su-re} further leads to
\begin{subequations}
\begin{eqnarray}
&&　p_n \wt{\bu}^{(i)}(a)=(p_n-\wt{a})\bu^{(i)}(\wt{a})+\bu^{(i+1)}(\wt{a})-\wt{S}^{(i,0)}(a,b)\bu^{(0)}(a). \label{eq:uia-dyna-a}
\end{eqnarray}
Similar relations to \eqref{eq:uia-dyna-a} also hold for the other lattice shifts, i.e., the
$\wh{\phantom{a}}-$ and $\wb{\phantom{a}}-$ shifts, associated with
the parameters $q_m$ and $r_h$, respectively:
\begin{eqnarray}
&& q_m \wh{\bu}^{(i)}(a)=(q_m-\wh{a})\bu^{(i)}(\wh{a})+\bu^{(i+1)}(\wh{a})-\wh{S}^{(i,0)}(a,b)\bu^{(0)}(a), \label{eq:uia-dyna-b} \\
&& r_h \wb{\bu}^{(i)}(a)=(r_h-\wb{a})\bu^{(i)}(\wb{a})+\bu^{(i+1)}(\wb{a})-\wb{S}^{(i,0)}(a,b)\bu^{(0)}(a). \label{eq:uia-dyna-c}
\end{eqnarray}
\label{uia-dyna}
\end{subequations}
The system \eqref{uia-dyna} gives the evolution of $\bu^{(i)}(a)$. The evolution of $\tu^{(j)}(b)$
can be obtained similarly, which are
\begin{subequations}
\begin{eqnarray}
&& p_n \tu^{(j)}(b)=(p_n+b)\wt{\tu}^{(j)}(\dt{b})-\wt{\tu}^{(j+1)}(\dt{b})+\wt{\tu}^{(0)}(b)S^{(0,j)}(a,b), \label{eq:tujb-dyna-a} \\
&& q_m \tu^{(j)}(b)=(q_m+b)\wh{\tu}^{(j)}(\dh{b})-\wh{\tu}^{(j+1)}(\dh{b})+\wh{\tu}^{(0)}(b)S^{(0,j)}(a,b), \label{eq:tujb-dyna-b} \\
&& r_h \tu^{(j)}(b)=(r_h+b)\wb{\tu}^{(j)}(\db{b})-\wb{\tu}^{(j+1)}(\db{b})+\wb{\tu}^{(0)}(b)S^{(0,j)}(a,b). \label{eq:tujb-dyna-c}
\end{eqnarray}
\label{tujb-dyna}
\end{subequations}
The evolution of $S^{(i,j)}(a,b)$ can be constructed from \eqref{uia-dyna} by left-multiplying
row vector $\ts (b\bI_{N'}+\bL)^j\bC$ and using \eqref{ce-2-b}, which reads
\begin{subequations}
\begin{eqnarray}
&& (p_n+b)\wt{S}^{(i,j)}(a,\ut{b})-\wt{S}^{(i,j+1)}(a,\ut{b}) \nn \\
&& ~~~~~~~~~~~~~~~~~~~~~~~~=(p_n-\wt{a})S^{(i,j)}(\wt{a},b)
 +S^{(i+1,j)}(\wt{a},b)-\wt{S}^{(i,0)}(a,b)S^{(0,j)}(a,b), \label{eq:S-dyna-a} \\
&& (q_m+b)\wh{S}^{(i,j)}(a,\dh{b})-\wh{S}^{(i,j+1)}(a,\dh{b}) \nn \\
&& ~~~~~~~~~~~~~~~~~~~~~~~~=(q_m-\wh{a})S^{(i,j)}(\wh{a},b)
 +S^{(i+1,j)}(\wh{a},b)-\wh{S}^{(i,0)}(a,b)S^{(0,j)}(a,b), \label{eq:S-dyna-b} \\
&& (r_h+b)\wb{S}^{(i,j)}(a,\db{b})-\wb{S}^{(i,j+1)}(a,\db{b}) \nn \\
&& ~~~~~~~~~~~~~~~~~~~~~~~~=(r_h-\wb{a})S^{(i,j)}(\wb{a},b)
 +S^{(i+1,j)}(\wb{a},b)-\wb{S}^{(i,0)}(a,b)S^{(0,j)}(a,b). \label{eq:S-dyna-c}
\end{eqnarray}
\label{S-dyna}
\end{subequations}
System \eqref{S-dyna} can also be derived from \eqref{eq:tujb-dyna-a}, \eqref{eq:tujb-dyna-b} and
\eqref{eq:tujb-dyna-c} by, respectively, right-multiplying
column vectors $\bC(\wt{a}\bI_{N}+\bK)^i\wt{\br}$, $\bC(\wh{a}\bI_{N}+\bK)^i\wh{\br}$ and $\bC(\wb{a}\bI_{N}+\bK)^i\wb{\br}$.

\subsection{Closed form lattice equations} \label{Va-NlKP}

From relation \eqref{S-dyna}, several non-autonomous lattice equations can be constructed.
To get them we introduce the following variables
\begin{subequations}
\begin{align}
& S^{(i,j)}=S^{(i,j)}(0,0),~~u=S^{(0,0)},~~v=1-S^{(-1,0)},~~w=1-S^{(0,-1)}, \label{uvw-def} \\
& v_a=1-S^{(-1,0)}(a,0),~~w_b=1-S^{(0,-1)}(0,b),~~s_{a,b}=S^{(-1,-1)}(a,b), \label{vawb-def} \\
& s_a=a-S^{(-1,1)}(a,0),~~t_b=S^{(1,-1)}(0,b)-b.
\end{align}
\label{vari}
\end{subequations}
Following the autonomous case \cite{N-KP,WZ}, we will see in the following non-autonomous lpKP
equation with the dependent variable $u$, non-autonomous lpmKP equation with the dependent variable $v$ or $w$,
non-autonomous lKP-NQC equation with the dependent variable $s_{a,b}$ where $a,b$ are constants,
and non-autonomous lSKP equation with a new dependent variable $z$, which is related to $S^{(-1,-1)}$
through a simple transformation. The non-autonomous asymmetric lpmKP equation appears
with dependent variables $v_{p_{n-1}}$ (or $v_{q_{m-1}}$ or $v_{r_{h-1}}$) or $w_{-p_n}$ (or $w_{-q_m}$ or $w_{-r_h}$).

\subsubsection{Non-autonomous lpKP equation}

When $i=j=0$ and $a=b=0$, \eqref{S-dyna} gives rise to
\begin{subequations}
\begin{eqnarray}
& p_n\wt{u}-\wt{S}^{(0,1)}=p_n u+S^{(1,0)}-\wt{u}u, \label{u-dyna-a} \\
& q_m\wh{u}-\wh{S}^{(0,1)}=q_m u+S^{(1,0)}-\wh{u}u, \label{u-dyna-b} \\
& r_h\wb{u}-\wb{S}^{(0,1)}=r_h u+S^{(1,0)}-\wb{u}u. \label{u-dyna-c}
\end{eqnarray}
\label{u-dyna}
\end{subequations}
Eliminating $S^{(0,1)}$ and $S^{(1,0)}$ in \eqref{u-dyna}, we get the non-autonomous lpKP equation
\begin{eqnarray}
(p_n-\wt{u})(q_m-r_h+\wt{\wb{u}}-\wh{\wt{u}})
+(q_m-\wh{u})(r_h-p_n+\wh{\wt{u}}-\wh{\wb{u}})+
(r_h-\wb{u})(p_n-q_m+\wh{\wb{u}}-\wt{\wb{u}})=0. \label{LPKP}
\end{eqnarray}
By recombining \eqref{LPKP}, two equivalent forms can be derived as
\begin{eqnarray}
(p_n+\wh{\wb{u}})(q_m-r_h+\wb{u}-\wh{u})
+(q_m+\wt{\wb{u}})(r_h-p_n+\wt{u}-\wb{u})+
(r_h+\wh{\wt{u}})(p_n-q_m+\wh{u}-\wt{u})=0,
\label{LPKP-1}
\end{eqnarray}
and
\begin{eqnarray}
\frac{(p_n-r_h+\wb{u}-\wt{u})^{\wh{\phantom{a}}}}{p_n-r_h+\wb{u}-\wt{u}}=
\frac{(q_m-r_h+\wb{u}-\wh{u})^{\wt{\phantom{a}}}}{q_m-r_h+\wb{u}-\wh{u}}=
\frac{(p_n-q_m+\wh{u}-\wt{u})^{\wb{\phantom{a}}}}{p_n-q_m+\wh{u}-\wt{u}}.
\label{LPKP-2}
\end{eqnarray}

\subsubsection{Non-autonomous lpmKP equation and asymmetric equation}

According to two different values $i=-1,j=0$ and $i=0,j=-1$, two non-autonomous lpmKP equations can be
constructed, corresponding to dependent variables $v$ and $w$, respectively. In the following, we will directly
consider the evolutions of $v_a$ and $w_b$. The advantage is that one can construct the
non-autonomous asymmetric lpmKP equation easily. The two non-autonomous lpmKP equations will be derived
from equations of $v_a$ and $w_b$ by taking $a=b=0$.

\vspace{.3cm}

\noindent
{\bf \underline{Equation-I}:} \\

Taking $i=-1,~j=0$ and $b=0$ in \eqref{S-dyna} and noting that definition \eqref{vari}, we have
\begin{subequations}
\begin{eqnarray}
&&　\wt{s}_a=(p_n+u)\wt{v}_a-(p_n-\wt{a})v_{\wt{a}}, \label{eq:va-dyna-a} \\
&&　\wh{s}_a=(q_m+u)\wh{v}_a-(q_m-\wh{a})v_{\wh{a}}, \label{eq:va-dyna-b} \\
&&　\wb{s}_a=(r_h+u)\wb{v}_a-(r_h-\wb{a})v_{\wb{a}}. \label{eq:va-dyna-c}
\end{eqnarray}
\label{va-dyna}
\end{subequations}
By removing $s_a$, we obtain the following system
\begin{subequations}
\begin{eqnarray}
&& p_n-q_m+\wh{u}-\wt{u}=\frac{(p_n-\wh{\wt{a}})\wh{v}_{\wt{a}}-(q_m-\wh{\wt{a}})\wt{v}_{\wh{a}}}{\wh{\wt{v}}_a}, \label{eq:Mu1-a} \\
&& r_h-p_n+\wt{u}-\wb{u}=\frac{(r_h-\wb{\wt{a}})\wt{v}_{\wb{a}}-(p_n-\wb{\wt{a}})\wb{v}_{\wt{a}}}{\wt{\wb{v}}_a}, \label{eq:Mu1-b} \\
&& q_m-r_h+\wb{u}-\wh{u}=\frac{(q_m-\wb{\wh{a}})\wb{v}_{\wh{a}}-(r_h-\wb{\wh{a}})\wh{v}_{\wb{a}}}{\wh{\wb{v}}_a}. \label{eq:Mu1-c}
\end{eqnarray}
\label{Mu1}
\end{subequations}
Adding the above three relations, we get
\begin{eqnarray}
\frac{(p_n-\wh{\wt{a}})\wh{v}_{\wt{a}}-(q_m-\wh{\wt{a}})\wt{v}_{\wh{a}}}{\wh{\wt{v}}_a}
+\frac{(r_h-\wb{\wt{a}})\wt{v}_{\wb{a}}-(p_n-\wb{\wt{a}})\wb{v}_{\wt{a}}}{\wt{\wb{v}}_a}
+\frac{(q_m-\wb{\wh{a}})\wb{v}_{\wh{a}}-(r_h-\wb{\wh{a}})\wh{v}_{\wb{a}}}{\wh{\wb{v}}_a}=0.
\label{lmKP-va-a}
\end{eqnarray}
Substituting \eqref{Mu1} into \eqref{LPKP-2}, we can obtain an equivalent form of \eqref{lmKP-va-a}.
Note that when $a=0$, \eqref{lmKP-va-a} yields the non-autonomous lpmKP equation, i.e.
\begin{eqnarray}
\frac{p_n\wh{v}-q_m\wt{v}}{\wh{\wt{v}}}
+\frac{r_h\wt{v}-p_n\wb{v}}{\wt{\wb{v}}}
+\frac{q_m\wb{v}-r_h\wh{v}}{\wh{\wb{v}}}=0.
\label{lmKP-v}
\end{eqnarray}
In this case, system \eqref{Mu1} with $a=0$ supplies the Miura
transformation between non-autonomous lpKP equation \eqref{LPKP}
and non-autonomous lpmKP equation \eqref{lmKP-v}.

It is remarkable to note that the non-autonomous asymmetric lpmKP equation can be obtained from
\eqref{lmKP-va-a} by setting $a=p_{n-1}$, which is given by
\begin{eqnarray}
\frac{(p_n-q_m)\wt{v}_{p_{n-1}}}{\wh{\wt{v}}_{p_{n-1}}}
+\frac{(r_h-p_n)\wt{v}_{p_{n-1}}}{\wt{\wb{v}}_{p_{n-1}}}
+\frac{(q_m-p_{n-1})\wb{v}_{p_{n-1}}-(r_h-p_{n-1})\wh{v}_{p_{n-1}}}{\wh{\wb{v}}_{p_{n-1}}}=0,
\label{A-lmKP-p}
\end{eqnarray}
where $v_{p_{n-1}}$ is the dependent variable. In a similar fashion, by
assuming $a=q_{m-1}$ or $a=r_{h-1}$, the corresponding non-autonomous asymmetric lpmKP equation
can be derived
\begin{eqnarray}
\frac{(p_n-q_m)\wh{v}_{q_{m-1}}}{\wh{\wt{v}}_{q_{m-1}}}
+\frac{(r_h-q_{m-1})\wt{v}_{q_{m-1}}-(p_n-q_{m-1})\wb{v}_{q_{m-1}}}{\wt{\wb{v}}_{q_{m-1}}}
+\frac{(q_m-r_h)\wh{v}_{q_{m-1}}}{\wh{\wb{v}}_{q_{m-1}}}=0,
\label{A-lmKP-q}
\end{eqnarray}
or
\begin{eqnarray}
\frac{(p_n-r_{h-1})\wh{v}_{r_{h-1}}-(q_m-r_{h-1})\wt{v}_{r_{h-1}}}{\wh{\wt{v}}_{r_{h-1}}}
+\frac{(r_h-p_n)\wb{v}_{r_{h-1}}}{\wt{\wb{v}}_{r_{h-1}}}
+\frac{(q_m-r_{h})\wb{v}_{r_{h-1}}}{\wh{\wb{v}}_{r_{h-1}}}=0.
\label{A-lmKP-r}
\end{eqnarray}

\vspace{.3cm}

\noindent
{\bf \underline{Equation-II}:} \\

Analogue to the earlier analysis, system \eqref{S-dyna} with $i=0,j=-1$ and $a=0$ leads to
\begin{subequations}
\begin{eqnarray}
&&　t_b=-(p_n+b)\wt{w}_{\dt{b}}+(p_n-\wt{u})w_b, \label{eq:wb-dyna-a} \\
&&　t_b=-(q_m+b)\wh{w}_{\dh{b}}+(q_m-\wh{u})w_b, \label{eq:wb-dyna-b} \\
&&　t_b=-(r_h+b)\wb{w}_{\db{b}}+(r_h-\wb{u})w_b. \label{eq:wb-dyna-c}
\end{eqnarray}
\label{eq:wb-dyna}
\end{subequations}
Comparing the three relations in \eqref{eq:wb-dyna}, we get
\begin{subequations}
\begin{eqnarray}
&& p_n-q_m+\wh{u}-\wt{u}=\frac{(p_n+b)\wt{w}_{\dt{b}}-(q_m+b)\wh{w}_{\dh{b}}}{w_b}, \label{eq:Mu2-a} \\
&& r_h-p_n+\wt{u}-\wb{u}=\frac{(r_h+b)\wb{w}_{\db{b}}-(p_n+b)\wt{w}_{\dt{b}}}{w_b}, \label{eq:Mu2-b} \\
&& q_m-r_h+\wb{u}-\wh{u}=\frac{(q_m+b)\wh{w}_{\dh{b}}-(r_h+b)\wb{w}_{\db{b}}}{w_b}. \label{eq:Mu2-c}
\end{eqnarray}
\label{Mu2}
\end{subequations}
Taking \eqref{Mu2} into \eqref{LPKP-2} gives rise to
\begin{eqnarray}
 \frac{1}{\wh{w}_b}\frac{((p_n+b)\wt{w}_{\dt{b}}-(r_h+b)\wb{w}_{\db{b}}){\wh{\phantom{a}}}}{(p_n+b)\wt{w}_{\dt{b}}-(r_h+b)\wb{w}_{\db{b}}}
&=&\frac{1}{\wt{w}_b}\frac{((q_m+b)\wh{w}_{\dh{b}}-(r_h+b)\wb{w}_{\db{b}}){\wt{\phantom{a}}}}{(q_m+b)\wh{w}_{\dh{b}}-(r_h+b)\wb{w}_{\db{b}}} \nn \\
&=&\frac{1}{\wb{w}_b}\frac{((p_n+b)\wt{w}_{\dt{b}}-(q_m+b)\wh{w}_{\dh{b}}){\wb{\phantom{a}}}}{(p_n+b)\wt{w}_{\dt{b}}-(q_m+b)\wh{w}_{\dh{b}}}.
 \label{lmKP-wb-a}
\end{eqnarray}
When $b=0$, \eqref{lmKP-wb-a} leads to one more non-autonomous lpmKP equation 
\begin{eqnarray}
\frac{1}{\wh{w}}\frac{(p_n\wt{w}-r_h\wb{w}){\wh{\phantom{a}}}}{p_n\wt{w}-r_h\wb{w}}
=\frac{1}{\wt{w}}\frac{(q_m\wh{w}-r_h\wb{w}){\wt{\phantom{a}}}}{q_m\wh{w}-r_h\wb{w}}
=\frac{1}{\wb{w}}\frac{(p_n\wt{w}-q_m\wh{w}){\wb{\phantom{a}}}}{p_n\wt{w}-q_m\wh{w}}.
 \label{lmKP-w}
\end{eqnarray}
Likewise, system \eqref{Mu2} with $b=0$ provides the Miura transformation between non-autonomous lpKP equation \eqref{LPKP}
and non-autonomous lpmKP equation \eqref{lmKP-w}. If we suppose $b=-p_{n}$ or $b=-q_{m}$ or $b=-r_{h}$,
then the non-autonomous asymmetric lpmKP equation can also be described as
\begin{eqnarray}
\frac{\wb{\wh{w}}_{-p_{n}}\wt{w}_{-p_{n}}}{\wh{w}_{-p_{n}}\wb{w}_{-p_{n}}}
=\frac{((q_m-p_n)\wh{w}_{-p_{n}}-(r_h-p_n)\wb{w}_{-p_{n}}){\wt{\phantom{a}}}}{(q_m-p_n)\wh{w}_{-p_{n}}-(r_h-p_n)\wb{w}_{-p_{n}}},
\label{A-lmKP-wp}
\end{eqnarray}
or
\begin{eqnarray}
\frac{\wb{\wt{w}}_{-q_{m}}\wh{w}_{-q_{m}}}{\wt{w}_{-q_{m}}\wb{w}_{-q_{m}}}
=\frac{((p_n-q_m)\wt{w}_{-q_{m}}-(r_h-q_m)\wb{w}_{-q_{m}}){\wh{\phantom{a}}}}{(p_n-q_m)\wt{w}_{-q_{m}}-(r_h-q_m)\wb{w}_{-q_{m}}},
\label{A-lmKP-wq}
\end{eqnarray}
or
\begin{eqnarray}
\frac{\wh{\wt{w}}_{-r_{h}}\wb{w}_{-r_{h}}}{\wt{w}_{-r_{h}}\wh{w}_{-r_{h}}}
=\frac{((p_n-r_h)\wt{w}_{-r_{h}}-(q_m-r_h)\wh{w}_{-r_{h}}){\wb{\phantom{a}}}}{(p_n-r_h)\wt{w}_{-r_{h}}-(q_m-r_h)\wh{w}_{-r_{h}}}.
\label{A-lmKP-wr}
\end{eqnarray}

\subsubsection{Non-autonomous lKP-NQC equation and lSKP equation }

Let us examine the equation related to $s_{a,b}$. It is easy to know that
\eqref{S-dyna} with $i=j=-1$ gives rise to relation
\begin{subequations}
\begin{eqnarray}
&&　1+(p_n-\wt{a})s_{\wt{a},b}-(p_n+b)\wt{s}_{a,\dt{b}}=\wt{v}_aw_b, \label{sab-dyna-a} \\
&&　1+(q_m-\wh{a})s_{\wh{a},b}-(q_m+b)\wh{s}_{a,\dh{b}}=\wh{v}_aw_b, \label{sab-dyna-b} \\
&&　1+(r_h-\wb{a})s_{\wb{a},b}-(r_h+b)\wb{s}_{a,\db{b}}=\wb{v}_aw_b. \label{sab-dyna-c}
\end{eqnarray}
\label{sab-dyna}
\end{subequations}
By the identity
\begin{eqnarray}
\frac{(\wt{v}_a w_b)^{\wh{\phantom{a}}}}{(\wt{v}_aw_b)^{\wb{\phantom{a}}}}
=\frac{(\wh{v}_aw_b)^{\wt{\phantom{a}}}}{(\wb{v}_aw_b)^{\wt{\phantom{a}}}}\times
\frac{(\wb{v}_aw_b)^{\wh{\phantom{a}}}}{(\wh{v}_aw_b)^{\wb{\phantom{a}}}},
\end{eqnarray}
we arrive at
\begin{align}
\frac{(1+(p_n-\wt{a})s_{\wt{a},b}-(p_n+b)\wt{s}_{a,\dt{b}})^{\wh{\phantom{a}}}}
{(1+(p_n-\wt{a})s_{\wt{a},b}-(p_n+b)\wt{s}_{a,\dt{b}})^{\wb{\phantom{a}}}}
=& \frac{(1+(q_m-\wh{a})s_{\wh{a},b}-(q_m+b)\wh{s}_{a,\dh{b}})^{\wt{\phantom{a}}}}
{(1+(r_h-\wb{a})s_{\wb{a},b}-(r_h+b)\wb{s}_{a,\db{b}})^{\wt{\phantom{a}}}} \nn \\
&\times \frac{(1+(r_h-\wb{a})s_{\wb{a},b}-(r_h+b)\wb{s}_{a,\db{b}})^{\wh{\phantom{a}}}}
{(1+(q_m-\wh{a})s_{\wh{a},b}-(q_m+b)\wh{s}_{a,\dh{b}})^{\wb{\phantom{a}}}}.
\label{NQC-a}
\end{align}
When $a,b$ are complex constants, we can reduce \eqref{NQC-a} to the so-called non-autonomous lKP-NQC equation, i.e.
\begin{align}
\frac{(1+(p_n-a)s_{a,b}-(p_n+b)\wt{s}_{a,b})^{\wh{\phantom{a}}}}{(1+(p_n-a)s_{a,b}-(p_n+b)\wt{s}_{a,b})^{\wb{\phantom{a}}}}
=& \frac{(1+(q_m-a)s_{a,b}-(q_m+b)\wh{s}_{a,b})^{\wt{\phantom{a}}}}{(1+(r_h-a)s_{a,b}-(r_h+b)\wb{s}_{a,b})^{\wt{\phantom{a}}}} \nn \\
&\times \frac{(1+(r_h-a)s_{a,b}-(r_h+b)\wb{s}_{a,b})^{\wh{\phantom{a}}}}{(1+(q_m-a)s_{a,b}-(q_m+b)\wh{s}_{a,b})^{\wb{\phantom{a}}}}.
\label{NQC}
\end{align}
By setting $a=b=0$ and defining
\begin{eqnarray}
z=S^{(-1,-1)}-\bigg(\sum_{i=n_0}^{n-1}\frac{1}{p_i}+\sum_{j=m_0}^{m-1}\frac{1}{q_j}
+\sum_{l=h_0}^{h-1}\frac{1}{h_l}+z_0\bigg),~~z_0\in\mathbb{C}, \label{z-def}
\end{eqnarray}
\eqref{NQC} leads to
\begin{eqnarray}
\frac{(\wh{z}-\wh{\wt{z}})(\wt{z}-\wt{\wb{z}})
(\wb{z}-\wh{\wb{z}})}{(\wt{z}-\wh{\wt{z}})(\wb{z}-\wt{\wb{z}})
(\wh{z}-\wh{\wb{z}})}=1, \label{lSKP}
\end{eqnarray}
which is the non-autonomous lSKP equation.

It is worthy noting that \eqref{NQC-a} is a general equation,
from which all the rest equations obtained above can be
deduced by a direct choice of parameters.

\subsection{Non-autonomous blKP equation}\label{N-blKP}

To construct the non-autonomous blKP equation, we consider the following $\tau$-function
\begin{eqnarray}
\tau=|\bI_N+\bM\bC|=|\bI_{N'}+\bC\bM|. \label{tau}
\end{eqnarray}
The latter identity is a consequence of the general Weinstein-Aronszajn formula.
Now we discuss the evolution of function $\tau$. Taking $\wt{\phantom{a}}$-shift on \eqref{tau}
and noting that \eqref{M-dyna-a} and \eqref{ce-2-b}, we have
\begin{align}
\wt{\tau}& =|\bI_N+\wt{\bM}\bC|=|\bI_N+\bM\bC||\bI_N+\frac{1}{p_n}(\bI_N+\bM\bC)^{-1}\br\wt{\ts}\bC| \nn \\
&=\tau(1+\frac{1}{p_n}\wt{\ts}\bC(\bI_N+\bM\bC)^{-1}\br)\nn \\
&=\tau(1-\ts(-p_n\bI_{N'}+\bL)^{-1}\bC(\bI_N+\bM\bC)^{-1}\br)\nn \\
&=\tau w_{-p_n}. \label{tau-w}
\end{align}
 In a similar way, taking $\ut{\phantom{a}}$-shift on \eqref{tau}
and noting that \eqref{M-dyna-a} and \eqref{ce-2-a}, we can get $\ut{\tau}=\tau v_{p_{n-1}}$. Therefore, we find
\begin{align}
\frac{\wt{\tau}}{\tau}=w_{-p_n}=\frac{1}{\wt{v}_{p_{n-1}}}. \label{tau-vw}
\end{align}
Based on this relation, \eqref{eq:Mu1-a} with $a=p_{n-1}$ or \eqref{eq:Mu2-a} with $b=-p_{n}$ yields
\begin{subequations}
\begin{eqnarray}
&& p_n-q_m+\wh{u}-\wt{u}=(p_n-q_m)\frac{\wh{\wt{\tau}}\tau}{\wt{\tau}\wh{\tau}}. \label{u-tau-a}
\end{eqnarray}
In terms of the symmetric property of $(p_n,\wt{\phantom{a}})$, $(q_m,\wh{\phantom{a}})$ and $(r_h,\wb{\phantom{a}})$, we also get
\begin{eqnarray}
&& q_m-r_h+\wb{u}-\wh{u}=(q_m-r_h)\frac{\wh{\wb{\tau}}\tau}{\wb{\tau}\wh{\tau}}, \label{u-tau-b} \\
&& r_h-p_n+\wt{u}-\wb{u}=(r_h-p_n)\frac{\wb{\wt{\tau}}\tau}{\wb{\tau}\wt{\tau}}. \label{u-tau-c}
\end{eqnarray}
\label{u-tau}
\end{subequations}
System \eqref{u-tau} gives rise to the non-autonomous blKP equation, which reads
\begin{eqnarray}
(p_n-q_m)\wh{\wt{\tau}}\wb{\tau}+(q_m-r_h)\wh{\wb{\tau}}\wt{\tau}+(r_h-p_n)\wb{\wt{\tau}}\wh{\tau}=0.
\label{blKP}
\end{eqnarray}
Obviously, \eqref{u-tau} can be referred to as Miura transformation between non-autonomous lpKP equation \eqref{LPKP}
and non-autonomous blKP equation \eqref{blKP}. In terms of relation \eqref{tau-vw} together with its $(q_m,\wh{\phantom{a}})$
and $(r_h,\wb{\phantom{a}})$ counterparts, we know that equation \eqref{blKP} can be understood as the potential equation for
the non-autonomous asymmetric lpmKP equation.

Up to now, we have constructed all the non-autonomous lattice KP-type equations by using
generalized Cauchy matrix approach. In the next part, we will identify these non-autonomous 
lattice equations by introducing simple point transformations.

\subsection{Deformation}\label{Def}

In this subsection, we discuss the non-autonomous lattice KP-type
equations obtained in subsections \ref{Va-NlKP} and \ref{N-blKP} by means of simple point transformations,
where lattice parameters are absorbed in the deformed equations. In order to compare the non-autonomous case with the
autonomous case, we give the autonomous lattice KP-type equations in Appendix \ref{A-KP-Case}.
We will find that both of these two types of equation can be described in the same form. The form is formal,
which not means these two types of equations can be converted with each other.

\vspace{.3cm}

\noindent
{\bf \underline{Non-autonomous lpKP equation }:} \\

Through transformation
\begin{eqnarray}
u=x+\bigg(\sum_{i=n_0}^{n-1}p_i+\sum_{j=m_0}^{m-1}q_j
+\sum_{l=h_0}^{h-1}r_l+x_0\bigg),~~x_0\in\mathbb{C},
\end{eqnarray}
equation \eqref{LPKP} yields
\begin{eqnarray}
\wt{x}(\wh{\wt{x}}-\wt{\wb{x}})+\wh{x}(\wh{\wb{x}}-\wh{\wt{x}})+\wb{x}(\wt{\wb{x}}-\wh{\wb{x}})=0.
\label{eq:LPKP-x}
\end{eqnarray}

\vspace{.3cm}

\noindent
{\bf \underline{Non-autonomous lpmKP equation and asymmetric equation}:} \\

For non-autonomous lpmKP equation \eqref{lmKP-v}, we consider
point transformation
\begin{eqnarray}
v=y \bigg(\prod_{i=n_0}^{n-1}p_i\bigg)\bigg(\prod_{j=m_0}^{m-1}q_j\bigg)
\bigg(\prod_{l=h_0}^{h-1}r_l\bigg)~y_0,~~y_0\in\mathbb{C},
\label{v-y}
\end{eqnarray}
under which equation \eqref{lmKP-v} yields
\begin{eqnarray}
\frac{\wh{y}-\wt{y}}{\wh{\wt{y}}}+\frac{\wt{y}-\wb{y}}{\wb{\wt{y}}}+\frac{\wb{y}-\wh{y}}{\wh{\wb{y}}}=0.
\label{lpmKP-y}
\end{eqnarray}
Similarly, for non-autonomous lpmKP equation \eqref{lmKP-w}, we have
\begin{eqnarray}
\frac{1}{\wh{y'}}\frac{(\wt{y'}-\wb{y'}){\wh{\phantom{a}}}}{\wt{y'}-\wb{y'}}
=\frac{1}{\wt{y'}}\frac{(\wh{y'}-\wb{y'}){\wt{\phantom{a}}}}{\wh{y'}-\wb{y'}}
=\frac{1}{\wb{y'}}\frac{(\wt{y'}-\wh{y'}){\wb{\phantom{a}}}}{\wt{y'}-\wh{y'}},
 \label{lmKP-y'}
\end{eqnarray}
where variable $y'$ is defined by
\begin{eqnarray}
w=y' \bigg(\prod_{i=n_0}^{n-1}\frac{1}{p_i}\bigg)\bigg(\prod_{j=m_0}^{m-1}\frac{1}{q_j}\bigg)
\bigg(\prod_{l=h_0}^{h-1}\frac{1}{r_l}\bigg)~y'_0,~~y'_0\in\mathbb{C}.
\label{w-y'}
\end{eqnarray}

The transformations
\begin{eqnarray}
v_{p_{n-1}}=\xi\bigg(\prod_{j=m_0}^{m-1}(p_{n-1}-q_j)\bigg)\bigg(\prod_{l=h_0}^{h-1}(r_l-p_{n-1})\bigg)~\xi_0,~~\xi_0\in\mathbb{C},
\label{A-lmKP-tran}
\end{eqnarray}
and
\begin{eqnarray}
w_{-p_{n}}=\eta\bigg(\prod_{j=m_0}^{m-1}\frac{1}{(q_j-p_n)}\bigg)\bigg(\prod_{l=h_0}^{h-1}
\frac{1}{(r_l-p_n)}\bigg)~\eta_0,~~\eta_0\in\mathbb{C}
\label{A-lmKP-tran-1}
\end{eqnarray}
convert, respectively, the non-autonomous asymmetric lpmKP equations \eqref{A-lmKP-p} and \eqref{A-lmKP-wp} into
\begin{eqnarray}
\frac{\wt{\xi}}{\wh{\wt{\xi}}}
+\frac{\wt{\xi}}{\wb{\wt{\xi}}}-
\frac{\wh{\xi}+\wb{\xi}}{\wh{\wb{\xi}}}=0,
\label{A-lmKP-xi}
\end{eqnarray}
and
\begin{eqnarray}
\frac{\wb{\wh{\eta}}\wt{\eta}}{\wh{\eta}\wb{\eta}}=
\frac{\wh{\wt{\eta}}-\wb{\wt{\eta}}}{\wh{\eta}-\wb{\eta}}.
\label{A-lmKP-eta}
\end{eqnarray}
Similar analysis can also be done to equations \eqref{A-lmKP-q},
\eqref{A-lmKP-r}, \eqref{A-lmKP-wq} and \eqref{A-lmKP-wr}.

\vspace{.3cm}

\noindent
{\bf \underline{Non-autonomous lKP-NQC equation }:} \\

In the construction of non-autonomous lSKP equation \eqref{lSKP},
we assume $a=b=0$ in \eqref{NQC}. Now for complex constants $a,b$ with $a\neq 0$ or $b\neq0$,
we can make transformation
\begin{eqnarray}
s_{a,b}=z' \prod_{i=n_0}^{n-1}\bigg(\frac{p_i-a}{p_i+b}\bigg)\prod_{j=m_0}^{m-1}\bigg(\frac{q_j-a}{q_j+b}\bigg)
\prod_{l=h_0}^{h-1}\bigg(\frac{r_l-a}{r_l+b}\bigg)z'_0+\frac{1}{a+b},~~z'_0\in\mathbb{C},
\end{eqnarray}
under which non-autonomous lKP-NQC equation \eqref{NQC} yields
\begin{eqnarray}
\frac{(\wh{z'}-\wh{\wt{z'}})(\wt{z'}-\wt{\wb{z'}})
(\wb{z'}-\wh{\wb{z'}})}{(\wt{z'}-\wh{\wt{z'}})(\wb{z'}-\wt{\wb{z'}})
(\wh{z'}-\wh{\wb{z'}})}=1. \label{lSKP-z'}
\end{eqnarray}
It should be noted that this equation has the same form with \eqref{lSKP}.

\vspace{.3cm}

\noindent
{\bf \underline{Non-autonomous blKP equation }:} \\

For non-autonomous blKP equation \eqref{blKP}, we consider a new variable $\sigma$ defined by
\begin{eqnarray}
\sg=\tau\bigg(\prod_{i=n_0}^{n-1}\prod_{j=m_0}^{m-1}(p_i-q_j)\bigg)
\bigg(\prod_{j=m_0}^{m-1}\prod_{l=h_0}^{h-1}(q_j-r_l)\bigg)
\bigg(\prod_{l=h_0}^{h-1}\prod_{i=n_0}^{n-1}(r_l-p_i)\bigg)\sg_0,~~\sg_0\in\mathbb{C}.
\end{eqnarray}
Then by direct calculation we know \eqref{blKP} becomes
\begin{eqnarray}
\wh{\wt{\sg}}\wb{\sg}+\wh{\wb{\sg}}\wt{\sg}+\wb{\wt{\sg}}\wh{\sg}=0.
\label{blKP-f}
\end{eqnarray}

\section{Explicit solutions of DES \eqref{ce}}

According to the analysis of section \ref{GCMA}, we know that
all the non-autonomous lattice KP-type equations are given by scalar function
$S^{(i,j)}(a,b)=\ts(b\bI_{N'}+\bL)^j\bC(\bI_{N}+\bM\bC)^{-1}(a\bI_N+\bK)^i\br$ and $\tau$-function
$\tau=|\bI_{N}+\bM\bC|$, where
$\ts$, $\br$, $\bM$, $\bK$ and $\bL$ are defined by DES \eqref{ce}. Therefore, for deriving
exact solutions to the non-autonomous lattice KP-type equations, we just need to solve the
DES \eqref{ce}. In terms of the invariance of $S^{(i,j)}(a,b)$ ($\tau$-function is also invariant) and the covariance of the DES
\eqref{ce} under transformations \eqref{KL1-KL} and \eqref{Mrs-1} (see Subsec. \ref{ss-inva}), here we turn to
solve the canonical equation set
\begin{subequations}
\begin{eqnarray}
&& \Ga \bM+\bM \Lb= \br\ts, \label{nce-1} \\
&& p_n \wt{\br}=(p_n\bI_N+\Ga)\br,~q_m\wh{\br}=(q_m\bI_N+\Ga)\br,~r_h \wb{\br}=(r_h\bI_N+\Ga)\br,
 \label{nce-2}\\
&&\wt{\ts}(p_n\bI_{N'}-\Lb)=p_n\ts,~\wh{\ts}(q_m\bI_{N'}-\Lb)=q_m\ts,~ \wb{\ts}
(r_h\bI_{N'}-\Lb)=r_h\ts, \label{nce-3}
\end{eqnarray}
\label{nce}
\end{subequations}
where $\Ga$ and $\Lb$ are $N\times N$ and $N'\times N'$ matrices in canonical form, respectively.
In canonical DES \eqref{nce}, evolution equations \eqref{nce-2}
and \eqref{nce-3} are always used to determine discrete
plain wave factor vectors $\br,\ts$ and Sylvester equation \eqref{nce-1} is always used to define matrix $\bM$.
By the assumptions of $\bK$ and $\bL$ in \eqref{ce}, we know here $\mathcal{E}(\Ga)\bigcap \mathcal{E}(-\Lb)=\varnothing$,
$s\bI_N\pm \Ga$ and $s\bI_{N'}\pm \Lb$ are invertible for $s=0,p_n,q_m,r_h,a,b$.
In virtue of the canonical structures of both $\Ga$ and $\Lb$, it is possible to give a complete
classification for the solutions. The solving procedure of the Sylvester equation \eqref{nce-1}
has been given in detail in Ref.\cite{WZ},
where matrix $\bM$ was factorized as $\bM=\bF\bG\bH$ (see also continue case \cite{ZSF}). Here we just
list some main results. The most
general solution of the DES \eqref{nce} can be obtained when $\Ga$ and $\Lb$ are taken as (for notations, one can see
Appendix \ref{A:G3})
\begin{subequations}
\label{Ga,Lb-gen}
\begin{align}
& \Ga=\mathrm{Diag}\bigl(\Ga^{\tyb{N$_1$}}_{\ty{D}}(\{k_i\}^{N_1}_{1}),
\Ga^{\tyb{N$_2$}}_{\ty{J}}(k_{N_1+1}),\Ga^{\tyb{N$_3$}}_{\ty{J}}(k_{N_1+2}),\cdots,
\Ga^{\tyb{N$_s$}}_{\ty{J}}(k_{N_1+(s-1)})\bigr), \\
& \Lb=\mathrm{Diag}\bigl(\Lb^{\tyb{N$_1$'}}_{\ty{D}}(\{\ka_j\}^{N'_1}_{1}),
\Lb^{\tyb{N$_2$'}}_{\ty{J}}(\ka_{N'_1+1}),\Lb^{\tyb{N$_3$'}}_{\ty{J}}(\ka_{N'_1+2}),\cdots,
\Lb^{\tyb{N$_s$'}}_{\ty{J}}(\ka_{N'_1+(s-1)})\bigr).
\end{align}
\end{subequations}
Then from \eqref{nce} we have solutions
\begin{equation}
\br=\left(
\begin{array}{l}
\br_{\ty{D}}^{\tyb{N$_1$}}(\{k_i\}_{1}^{N_1})\\
\br_{\ty{J}}^{\tyb{N$_2$}}(k_{N_1+1})\\
\br_{\ty{J}}^{\tyb{N$_3$}}(k_{N_1+2})\\
\vdots\\
\br_{\ty{J}}^{\tyb{N$_s$}}(k_{N_1+(s-1)})
\end{array}
\right),~~~\ts=\left(
\begin{array}{l}
\ts_{\ty{D}}^{\tyb{N$_1$'}}(\{\ka_j\}_{1}^{N'_1})^T\\
\ts_{\ty{J}}^{\tyb{N$_2$'}}(\ka_{N'_1+1})^T\\
\ts_{\ty{J}}^{\tyb{N$_3$'}}(\ka_{N'_1+2})^T\\
\vdots\\
\ts_{\ty{J}}^{\tyb{N$_s$'}}(\ka_{N'_1+(s-1)})^T
\end{array}
\right)^T,
\end{equation}
and $\bM=\bF\bG \bH$, where
\begin{align}
&\bF=\mathrm{Diag}\bigl(
\Ga^{\tyb{N$_1$}}_{\ty{D}}(\{\ro_i\}^{N_1}_{1}),
\bT^{\tyb{N$_2$}}(k_{N_1+1}),\bT^{\tyb{N$_3$}}(k_{N_1+2}),\cdots,
\bT^{\tyb{N$_s$}}(k_{N_1+(s-1)})
\bigr),\label{KP-r-M-g-F}\\
&\bH=\mathrm{Diag}\bigl(
\Lb^{\tyb{N$_1$'}}_{\ty{D}}(\{\vr_j\}^{N'_1}_{1}),
\bH^{\tyb{N$_2$'}}(\ka_{N'_1+1}),
\bH^{\tyb{N$_3$'}}(\ka_{N'_1+2}),
\cdots,
\bH^{\tyb{N$_s$'}}(\ka_{N'_1+(s-1)})\bigr),\label{KP-r-M-g-H}
\end{align}
and $\bG$ possesses block structure
\begin{equation}
\bG=(\bG_{i,j})_{s\times s},
\label{KP-r-M-g-G1}
\end{equation}
with
\begin{subequations}\label{KP-r-M-g-G2}
\begin{align}
& \bG_{1,1}=\bG^{\tyb{N$_1$;N$_1$'}}_{\ty{DD}}(\{k_i\}^{N_1}_{1};\{\ka_j\}^{N'_1}_{1}), \label{KP-bs-sol-1}\\
& \bG_{1,j}=\bG^{\tyb{N$_1$;N$_j$'}}_{\ty{DJ}}(\{k_i\}^{N_1}_{1};\ka_{N'_1+j-1}),~~~(1<j\leq s), \label{KP-bs-sol-2}\\
& \bG_{i,1}=\bG^{\tyb{N$_i$;N$_1$'}}_{\ty{JD}}(k_{N_1+i-1};\{\ka_j\}^{N'_1}_{1}),~~~(1<i\leq s), \label{KP-bs-sol-3}\\
& \bG_{i,j}=\bG^{\tyb{N$_i$;N$_j$'}}_{\ty{JJ}}(k_{N_1+i-1};\ka_{N'_1+j-1}),~~~(1<i,j\leq s). \label{KP-bs-sol-4}
\end{align}
\end{subequations}
The corresponding solution is called mixed solution.

Some special solutions can be derived from the mixed solution by setting order. For example, when $N_1=N$ and $N'_1=N'$ with
$N_i=N'_j=0 (i,j=2,3,\ldots,s)$, the corresponding solution can be described as
\begin{align}
\br=\br_{\hbox{\tiny{\it D}}}^{\hbox{\tiny{[{\it N}]}}}(\{k_i\}_{1}^{N}),~~
\ts=\ts_{\hbox{\tiny{\it D}}}^{\hbox{\tiny{[{\it N'}]}}}(\{\ka_j\}_{1}^{N'}),
\end{align}
and
\begin{subequations}
\begin{equation}
\bM=\bF \bG \bH =\Bigl(\frac{\ro_i \vr_j}{k_i+\ka_j}\Bigr)_{N\times N'}\, ,
\end{equation}
where
\begin{equation}
\bF=\Ga^{\tyb{N}}_{\ty{D}}(\{\ro_i\}^{N}_{1}),~~
\bG=\bG^{\tyb{N;N'}}_{\ty{DD}}(\{k_i\}^{N}_{1};\{\ka_j\}^{N'}_{1}),~~
\bH=\Lb^{\tyb{N'}}_{\ty{D}}(\{\vr_j\}^{N'}_{1}).
\end{equation}
\end{subequations}
This leads to the multi-soliton solutions. When $N_2=N$ and $N'_2=N'$ with
$N_i=N'_j=0 (i,j=1,3,\ldots,s)$, the corresponding solution is given by
\begin{align}
\br=\br_{\hbox{\tiny{\it J}}}^{\hbox{\tiny{[{\it N}]}}}(k_{N_1+1}),~~\ts=\ts_{\hbox{\tiny{\it J}}}^{\hbox{\tiny{[{\it N'}]}}}(\ka_{N'_1+1}),
\end{align}
and
\begin{subequations}
\begin{equation}
\bM=\bF  \bG  \bH,
\end{equation}
where
\begin{equation}
\bF=\bT^{\tyb{N}}(k_{N_1+1}),~~\bG=\bG^{\tyb{N;N'}}_{\ty{JJ}}(k_{N_1+1};\ka_{N'_1+1}),~~ \bH=\bH^{\tyb{N'}}(\ka_{N'_1+1}).
\end{equation}
\end{subequations}
This is called Jordan-block solutions or multi-pole solutions, which can be derived from multi-soliton solutions through limit procedure.
Since we need $\Ga$ and $\Lb$ to satisfy invertible conditions,
eigenvalues of $\Ga$ and $\Lb$ can not be zero. Therefore here we can not obtain rational solutions.

\section{Lax representation}

The Lax pair for Hirota's DAGTE (Discrete analogue of a generalized Toda equation, which is
equivalent to autonomous blKP equation by independent variable transformation)
was firstly given by Hirota in Ref. \cite{H-BLKP}. Subsequently,
Wiersma and Capel \cite{N-KP-CL}
discussed the Lax representation of autonomous lpKP equation by DL method.
In this section, we will make use of a similar method in Ref. \cite{N-KP-CL} to
systematically construct the Lax representation of the non-autonomous lattice KP-type equations
with the help of evolutions \eqref{uia-dyna} and \eqref{tujb-dyna}.
For convenience, we introduce operators $T_i~(i=1,2,3)$, which are
defined by $T_1 f=\wt{f}$, $T_2 f=\wh{f}$ and $T_3 f=\wb{f}$. Their inverse are shown by
$T^{-1}_1 f=\dt{f}$, $T^{-1}_2 f=\dh{f}$ and $T^{-1}_3 f=\db{f}$. Notations $(\bu^{(i)}(a))_0$ and  $(\tu^{(j)}(b))_0$ indicate
the first component of vectors $\bu^{(i)}(a)$ and  $\tu^{(j)}(b)$, respectively.
In addition, we also denote $\bu^{(i)}(0)=\bu^{(i)}$ and $\tu^{(j)}(0)=\tu^{(j)}$.

\vspace{.3cm}

\noindent
{\bf \underline{Non-autonomous lpKP equation and blKP equation}:} \\

The first component of system \eqref{uia-dyna} with $i=0$ and $a=0$ yields
\begin{subequations}
\begin{eqnarray}
&& p_n T_1(\bu^{(0)})_0=p_n (\bu^{(0)})_0+(\bu^{(1)})_0-\wt{u}(\bu^{(0)})_0, \label{eq:u00-ld-a}\\
&& q_m T_2(\bu^{(0)})_0=q_m(\bu^{(0)})_0+(\bu^{(1)})_0-\wh{u}(\bu^{(0)})_0, \label{eq:u00-ld-b} \\
&& r_h T_3(\bu^{(0)})_0=r_h(\bu^{(0)})_0+(\bu^{(1)})_0-\wb{u}(\bu^{(0)})_0. \label{eq:uia-dyna-c}
\end{eqnarray}
\label{u00-ld}
\end{subequations}
Eliminating $(\bu^{(1)})_0$, we get a linear system
\begin{subequations}
\begin{eqnarray}
&& (p_nT_1-q_mT_2)(\bu^{(0)})_0=\phi_3(\bu^{(0)})_0,~~\phi_3=(p_n-q_m+\wh{u}-\wt{u}), \label{eq:u12-lp}\\
&& (q_mT_2-r_hT_3)(\bu^{(0)})_0=\phi_1(\bu^{(0)})_0,~~\phi_1=(q_m-r_h+\wb{u}-\wh{u}), \label{eq:u23-lp} \\
&& (r_hT_3-p_nT_1)(\bu^{(0)})_0=\phi_2(\bu^{(0)})_0,~~\phi_2=(r_h-p_n+\wt{u}-\wb{u}), \label{eq:u31-lp}
\end{eqnarray}
\label{u-lp}
\end{subequations}
where $(\bu^{(0)})_0$ can be referred to as the potential function. The compatibility condition of \eqref{u-lp} leads to
\begin{eqnarray}
\frac{T_1\phi_1}{\phi_1}=\frac{T_2\phi_2}{\phi_2}=\frac{T_3\phi_3}{\phi_3},
\label{lpKP-phi}
\end{eqnarray}
which is the non-autonomous lpKP equation \eqref{LPKP-2}. In fact,
$r_h*T_3 \eqref{eq:u12-lp}+p_n*T_1\eqref{eq:u23-lp}+q_m*T_2 \eqref{eq:u31-lp}$ gives rise to
\begin{eqnarray}
(T_3\phi_3)(r_hT_3(\bu^{(0)})_0)+(T_1\phi_1)(p_nT_1(\bu^{(0)})_0)+(T_2\phi_2)(q_mT_2(\bu^{(0)})_0)=0.
\label{lpKP-lp-com-1}
\end{eqnarray}
Removing $r_hT_3(\bu^{(0)})_0$ and $q_mT_2(\bu^{(0)})_0$ by \eqref{eq:u31-lp} and \eqref{eq:u12-lp}, we have
\begin{eqnarray}
p_n(T_1\phi_1+T_2\phi_2+T_3\phi_3)T_1(\bu^{(0)})_0+\big((T_3\phi_3)\phi_2-(T_2\phi_2)\phi_3)\big)(\bu^{(0)})_0=0.
\label{lpKP-lp-com-2}
\end{eqnarray}
Noting that $T_1\phi_1+T_2\phi_2+T_3\phi_3=0$, we get $(T_3\phi_3)\phi_2-(T_2\phi_2)\phi_3=0$, i.e.,
$\frac{T_2\phi_2}{\phi_2}=\frac{T_3\phi_3}{\phi_3}$.
In a similar fashion, if we remove $r_hT_3(\bu^{(0)})_0$ and $p_nT_1(\bu^{(0)})_0$ by \eqref{eq:u23-lp} and \eqref{eq:u12-lp}
or $p_nT_1(\bu^{(0)})_0$ and $q_mT_2(\bu^{(0)})_0$ by \eqref{eq:u31-lp} and \eqref{eq:u23-lp},
we can also have $\frac{T_1\phi_1}{\phi_1}=\frac{T_3\phi_3}{\phi_3}$ or $\frac{T_1\phi_1}{\phi_1}=\frac{T_2\phi_2}{\phi_2}$.
Thus \eqref{lpKP-phi} holds. This means linear system \eqref{u-lp} can be viewed as the Lax representation of the non-autonomous lpKP
equation.

Focusing on system \eqref{tujb-dyna} with $j=0$ and $b=0$, we can obtain another Lax representation for
the non-autonomous lpKP equation, which is described as
\begin{subequations}
\begin{eqnarray}
&& (p_nT_2-q_mT_1)(\tu^{(0)})_0=\phi_3T_1T_2(\tu^{(0)})_0, \label{eq:tu12-lp}\\
&& (q_mT_3-r_hT_2)(\tu^{(0)})_0=\phi_1T_2T_3(\tu^{(0)})_0, \label{eq:tu23-lp} \\
&& (r_hT_1-p_nT_3)(\tu^{(0)})_0=\phi_2T_1T_3(\tu^{(0)})_0, \label{eq:tu31-lp}
\end{eqnarray}
\label{tu-lp}
\end{subequations}
where $(\tu^{(0)})_0$ is the potential function and $\phi_i~(i=1,2,3)$ are defined by \eqref{u-lp}.

Thanks to the relations \eqref{Mu1}, \eqref{Mu2} and \eqref{u-tau},
$(\bu^{(0)})_0$ and $(\tu^{(0)})_0$ can always be
viewed as the potential functions to construct the Lax representations of non-autonomous lpmKP equation,
non-autonomous asymmetric lpmKP equation and non-autonomous blKP equation. For example, for the
non-autonomous blKP equation \eqref{blKP}, the corresponding Lax representations can be described as
\begin{subequations}
\begin{eqnarray}
&& (p_nT_1-q_mT_2)(\bu^{(0)})_0=\vi_3(\bu^{(0)})_0,~~\vi_3=(p_n-q_m)\frac{\wh{\wt{\tau}}\tau}{\wt{\tau}\wh{\tau}}, \label{eq:tau12-lp}\\
&& (q_mT_2-r_hT_3)(\bu^{(0)})_0=\vi_1(\bu^{(0)})_0,~~\vi_1=(q_m-r_h)\frac{\wh{\wb{\tau}}\tau}{\wb{\tau}\wh{\tau}}, \label{eq:tau23-lp} \\
&& (r_hT_3-p_nT_1)(\bu^{(0)})_0=\vi_2(\bu^{(0)})_0,~~\vi_2=(r_h-p_n)\frac{\wb{\wt{\tau}}\tau}{\wb{\tau}\wt{\tau}}, \label{eq:tau31-lp}
\end{eqnarray}
\label{tau-lp}
\end{subequations}
or
\begin{subequations}
\begin{eqnarray}
&& (p_nT_2-q_mT_1)(\tu^{(0)})_0=\vi_3T_1T_2(\tu^{(0)})_0, \label{eq:tau12-lp-t}\\
&& (q_mT_3-r_hT_2)(\tu^{(0)})_0=\vi_1T_2T_3(\tu^{(0)})_0, \label{eq:tau23-lp-t} \\
&& (r_hT_1-p_nT_3)(\tu^{(0)})_0=\vi_2T_1T_3(\tu^{(0)})_0.\label{eq:tau31-lp-t}
\end{eqnarray}
\label{tu-lp-t}
\end{subequations}

In the following, we will consider another Lax representations of
non-autonomous lpmKP equation and non-autonomous asymmetric lpmKP equation.
The proofs are similar to the non-autonomous lpKP case, which are omitted here.

\vspace{.3cm}

\noindent
{\bf \underline{Non-autonomous lpmKP-I equation and asymmetric equation}:} \\

We consider potential function $\big(\bu^{(-1)}/v\big)_0$. It is easy to know that
\eqref{uia-dyna} with $a=0$ and $i=-1$ implies
\begin{subequations}
\begin{eqnarray}
&& p_n T_1\bigg(\frac{\bu^{(-1)}}{v}\bigg)_0
=\frac{p_nv}{\wt{v}}\bigg(\frac{\bu^{(-1)}}{v}\bigg)_0+(\bu^{(0)})_0, \label{eq:u-1a-ld-a}\\
&& q_m T_2\bigg(\frac{\bu^{(-1)}}{v}\bigg)_0
=\frac{q_mv}{\wh{v}}\bigg(\frac{\bu^{(-1)}}{v}\bigg)_0+(\bu^{(0)})_0, \label{eq:u-1a-ld-b} \\
&& r_h T_3\bigg(\frac{\bu^{(-1)}}{v}\bigg)_0
=\frac{r_hv}{\wb{v}}\bigg(\frac{\bu^{(-1)}}{v}\bigg)_0+(\bu^{(0)})_0. \label{eq:u-1a-ld-c}
\end{eqnarray}
\label{uia-dyna-T}
\end{subequations}
By removing $(\bu^{(0)})_0$, we get a linear system
\begin{subequations}
\begin{eqnarray}
&& (p_nT_1-q_mT_2)\bigg(\frac{\bu^{(-1)}}{v}\bigg)_0=\var_3
\bigg(\frac{\bu^{(-1)}}{v}\bigg)_0,~~ \var_3=\bigg(\frac{p_n}{\wt{v}}-
\frac{q_m}{\wh{v}}\bigg)v, \label{eq:v12-ld}\\
&& (q_mT_2-r_hT_3)\bigg(\frac{\bu^{(-1)}}{v}\bigg)_0=\var_1
\bigg(\frac{\bu^{(-1)}}{v}\bigg)_0,~~\var_1=\bigg(\frac{q_m}{\wh{v}}-
\frac{r_h}{\wb{v}}\bigg)v, \label{eq:v23-ld} \\
&& (r_hT_3-p_nT_1)\bigg(\frac{\bu^{(-1)}}{v}\bigg)_0=\var_2
\bigg(\frac{\bu^{(-1)}}{v}\bigg)_0,~~\var_2=\bigg(\frac{r_h}{\wb{v}}
-\frac{p_n}{\wt{v}}\bigg)v,\label{eq:v31-ld}
\end{eqnarray}
\label{va-lp}
\end{subequations}
which is one more Lax representation of the non-autonomous lpmKP equation \eqref{lmKP-v}.

For non-autonomous asymmetric lpmKP equation \eqref{A-lmKP-p}, we can take $a=p_{n-1}$ and $i=-1$ in \eqref{uia-dyna} and
remove $(\bu^{(0)})_0$ and finally arrive at
\begin{subequations}
\begin{eqnarray}
&& (p_nT_1-q_mT_2)\bigg(\frac{\bu^{(-1)}(p_{n-1})}{v_{p_{n-1}}}\bigg)_0
=\psi_3\bigg(\frac{\bu^{(-1)}(p_{n-1})}{v_{p_{n-1}}}\bigg)_0,\\
&& (q_mT_2-r_hT_3)\bigg(\frac{\bu^{(-1)}(p_{n-1})}{v_{p_{n-1}}}\bigg)_0
=\psi_1\bigg(\frac{\bu^{(-1)}(p_{n-1})}{v_{p_{n-1}}}\bigg)_0, \\
&& (r_hT_3-p_nT_1)\bigg(\frac{\bu^{(-1)}(p_{n-1})}{v_{p_{n-1}}}\bigg)_0
=\psi_2\bigg(\frac{\bu^{(-1)}(p_{n-1})}{v_{p_{n-1}}}\bigg)_0,
\end{eqnarray}
\end{subequations}
where
\begin{eqnarray*}
\psi_1=\bigg(\frac{q_m-p_{n-1}}{\wh{v}_{p_{n-1}}}-\frac{r_h-p_{n-1}}{\wb{v}_{p_{n-1}}}\bigg)v_{p_{n-1}},~
\psi_2=\frac{(r_h-p_{n-1})v_{p_{n-1}}}{\wb{v}_{p_{n-1}}},~
\psi_3=-\frac{(q_m-p_{n-1})v_{p_{n-1}}}{\wh{v}_{p_{n-1}}},
\end{eqnarray*}
and $(\bu^{(-1)}(p_{n-1})/v_{p_{n-1}})_0$ is the potential function.

\vspace{.3cm}

\noindent
{\bf \underline{Non-autonomous lpmKP-II equation and asymmetric equation}:} \\

Similar as before, the Lax representation for the non-autonomous lpmKP-II equation \eqref{lmKP-w} can also be obtained by
taking $j=-1$ and $b=0$ in \eqref{tujb-dyna} and eliminating $(\tu^{(0)})_0$, which reads
\begin{subequations}
\begin{eqnarray}
&& (p_{n-1}T^{-1}_1-q_{m-1}T^{-1}_2)\bigg(\frac{\tu^{(-1)}}{w}\bigg)_0
=\ci_3\bigg(\frac{\tu^{(-1)}}{w}\bigg)_0,~~\ci_3=\bigg(\frac{p_{n-1}}{\ut{w}}-
\frac{q_{m-1}}{\dh{w}}\bigg)w, \\
&& (q_{m-1}T^{-1}_2-r_{h-1}T^{-1}_3)\bigg(\frac{\tu^{(-1)}}{w}\bigg)_0=
\ci_1\bigg(\frac{\tu^{(-1)}}{w}\bigg)_0,~~\ci_1=\bigg(\frac{q_{m-1}}{\dh{w}}-
\frac{r_{h-1}}{\db{w}}\bigg)w, \\
&& (r_{h-1}T^{-1}_3-p_{n-1}T^{-1}_1)\bigg(\frac{\tu^{(-1)}}{w}\bigg)_0=
\ci_2\bigg(\frac{\tu^{(-1)}}{w}\bigg)_0,~~\ci_2=\bigg(\frac{r_{h-1}}{\db{w}}
-\frac{p_{n-1}}{\ut{w}}\bigg)w,
\end{eqnarray}
\label{w-lp-2}
\end{subequations}
where $(\tu^{(-1)}/w)_0$ is the potential function.
Analogue as aforementioned, the Lax representation for the non-autonomous asymmetric lpmKP equation \eqref{A-lmKP-wp} can also be derived
\begin{subequations}
\begin{eqnarray}
&& (p_{n-1}T^{-1}_1-q_{m-1}T^{-1}_2)\bigg(\frac{\tu^{(-1)}(-p_n)}{w_{-p_n}}\bigg)_0
=\vs_3\bigg(\frac{\tu^{(-1)}(-p_n)}{w_{-p_n}}\bigg)_0, \\
&& (q_{m-1}T^{-1}_2-r_{h-1}T^{-1}_3)\bigg(\frac{\tu^{(-1)}(-p_n)}{w_{-p_n}}\bigg)_0=
\vs_1\bigg(\frac{\tu^{(-1)}(-p_n)}{w_{-p_n}}\bigg)_0, \\
&& (r_{h-1}T^{-1}_3-p_{n-1}T^{-1}_1)\bigg(\frac{\tu^{(-1)}(-p_n)}{w_{-p_n}}\bigg)_0=
\vs_2\bigg(\frac{\tu^{(-1)}(-p_n)}{w_{-p_n}}\bigg)_0,
\end{eqnarray}
\label{Aw-lp-2}
\end{subequations}
where
\begin{eqnarray*}
&& \vs_1=\bigg(\frac{q_{m-1}-p_n}{\dh{w}_{-p_n}}-\frac{r_{h-1}-p_n}{\db{w}_{-p_n}}\bigg)w_{-p_n},~
\vs_2=\frac{r_{h-1}-p_n}{\db{w}_{-p_n}}w_{-p_n}, ~\vs_3=-\frac{q_{m-1}-p_n}{\dh{w}_{-p_n}}w_{-p_n},
\end{eqnarray*}
and the potential function is $(\tu^{(-1)}(-p_n)/w_{-p_n})_0$.

%

\section*{Conclusion}

The study of non-autonomous discrete equation is always an interesting topic in discrete integrable system.
Comparing with autonomous discrete equation, non-autonomous discrete equation can be understood as an
equation with variable coefficients. The main
difference between autonomous case and non-autonomous case focus on the following aspects:
\begin{equation*}
\begin{array}{llcl}
\mathrm{spacing~parameters:}& (p,q,r) & \to& (p_n, q_m, r_h),\\
\mathrm{linear~ function:}& pn+qm+rh & \to& \sum^{n-1}_{i=n_0}p_i+\sum^{m-1}_{j=m_0}q_j+\sum^{h-1}_{l=h_0}r_l,\\
\mathrm{discrete~ exponential~ function:}&
(\frac{p+a}{p-b})^n(\frac{q+a}{q-b})^m(\frac{r+a}{r-b})^h & \to&
\prod_{i=n_0}^{n-1}(\frac{p_i+a}{p_i-b})\prod_{j=m_0}^{m-1}(\frac{q_j+a}{q_j-b})
\prod_{l=h_0}^{h-1}(\frac{r_l+a}{r_l-b}).
\end{array}
\end{equation*}
Since both of these two types of equations can be described in the same form through point transformation
(see also Subsec.\ref{Def} and Appendix \ref{A-KP-Case}), they should have many similar properties, 
such as multi-dimensional consistency.
A extreme important point of the non-autonomous discrete equation is that it is
usually used to reduce discrete Painlev\'{e} equation, which is always
the non-autonomous ordinary difference equation.

In this paper we investigate the non-autonomous lattice KP-type equations
in terms of generalized Cauchy matrix approach with the DES \eqref{ce} to be the starting point.
By introducing scalar function $S^{(i,j)}(a,b)=\ts(b\bI_{N'}+\bL)^j\bC(\bI_{N}+\bM\bC)^{-1}(a\bI_N+\bK)^i \br$
and considering its dynamical properties, several non-autonomous lattice KP-type equations are obtained, including
non-autonomous lpKP equation, non-autonomous lpmKP equation, non-autonomous asymmetric lpmKP equation,
non-autonomous lSKP equation and non-autonomous lKP-NQC equation. The non-autonomous blKP equation is derived by introducing
$\tau$-function $\tau=|\bI_{N}+\bM\bC|$. Through point transformations all the obtained equations are transformed into
simplified forms. For the purpose of finding exact solutions of the
DES \eqref{ce}, we consider the canonical forms $\Ga$ and $\Lb$ of matrices $\bK$ and $\bL$
and introduce transformation \eqref{Mrs-1}
to simplify it. As a result, the most general solution of the canonical DES is obtained when $\Ga$ and $\Lb$
are taken the general form (diagonal-Jordan block combination). The multi-soliton solutions and the multi-pole solutions
are also constructed as special cases. With the help of the dynamical properties of vector functions
$\bu^{(i)}(a)$ and $\tu^{(j)}(b)$, we construct Lax representations for the non-autonomous lattice KP-type equations. According
to different choice of the potential functions, several expressions have been revealed.

We hope the results given in present paper will be useful to study the discrete Painlev\'{e} equation.

\vskip 20pt \noindent{\bf Acknowledgments}~

This project is supported by the National Natural Science Foundation of China (Nos. 11301483, 11401529 and 11371323) and the
Natural Science Foundation of Zhejiang province (No. Y6100611).

\vskip 20pt

\begin{appendix}

\section{Autonomous discrete KP equations} \label{A-KP-Case}

The list of autonomous discrete KP equations are
\begin{subequations}\label{dkp-list}
\begin{align}
\mbox{lpKP equation}:
~~&(p-\wt{u})(q-r+\wt{\wb{u}}-\wh{\wt{u}})
+(q-\wh{u})(r-p+\wh{\wt{u}}-\wh{\wb{u}})\nn \\
& +(r-\wb{u})(p-q+\wh{\wb{u}}-\wt{\wb{u}})=0,\\
\mbox{lpmKP-I equation}:~~&\frac{p\wh{v}-q\wt{v}}{\wh{\wt{v}}}
+\frac{r\wt{v}-p\wb{v}}{\wt{\wb{v}}}
+\frac{q\wb{v}-r\wh{v}}{\wh{\wb{v}}}=0,\\
\mbox{lpmKP-II equation}:~~&\frac{1}{\wh{w}}\frac{(p\wt{w}-r\wb{w}){\wh{\phantom{a}}}}{p\wt{w}-r\wb{w}}
=\frac{1}{\wt{w}}\frac{(q\wh{w}-r\wb{w}){\wt{\phantom{a}}}}{q\wh{w}-r\wb{w}}
=\frac{1}{\wb{w}}\frac{(p\wt{w}-q\wh{w}){\wb{\phantom{a}}}}{p\wt{w}-q\wh{w}},\\
\mbox{asymmetric lpmKP-I equation}:~~&\frac{(p-q)\wt{v}_{p}}{\wh{\wt{v}}_{p}}
+\frac{(r-p)\wt{v}_{p}}{\wt{\wb{v}}_{p}}
+\frac{(q-p)\wb{v}_{p}-(r-p)\wh{v}_{p}}{\wh{\wb{v}}_{p}}=0, \\
\mbox{asymmetric lpmKP-II equation}:~~&
\frac{\wb{\wh{w}}_{-p}\wt{w}_{-p}}{\wh{w}_{-p}\wb{w}_{-p}}
=\frac{((q-p)\wh{w}_{-p}-(r-p)\wb{w}_{-p}){\wt{\phantom{a}}}}{(q-p)\wh{w}_{-p}-(r-p)\wb{w}_{-p}}, \\
\mbox{lKP-NQC equation}:~~&
\frac{(1+(p-a)s_{a,b}-(p+b)\wt{s}_{a,b})^{\wh{\phantom{a}}}}{(1+(p-a)s_{a,b}-(p+b)\wt{s}_{a,b})^{\wb{\phantom{a}}}} \nn \\ 
&= \frac{(1+(q-a)s_{a,b}-(q+b)\wh{s}_{a,b})^{\wt{\phantom{a}}}}{(1+(r-a)s_{a,b}-(r+b)\wb{s}_{a,b})^{\wt{\phantom{a}}}} \nn \\
&~~\times \frac{(1+(r-a)s_{a,b}-(r+b)\wb{s}_{a,b})^{\wh{\phantom{a}}}}{(1+(q-a)s_{a,b}-(q+b)\wh{s}_{a,b})^{\wb{\phantom{a}}}}, \\
\mbox{blKP equation}:~~&
(p-q)\wh{\wt{\tau}}\wb{\tau}+(q-r)\wh{\wb{\tau}}\wt{\tau}+(r-p)\wb{\wt{\tau}}\wh{\tau}=0.
\end{align}
\end{subequations}
where lattice parameters $p,q,r$ are complex constants.

For equations in list \eqref{dkp-list}, we consider transformations
\begin{subequations}\label{dkp-trans}
\begin{align}
& u=x+pn+qm+rh+x_0,\\
& v=yp^nq^mr^hy_0, \\
& w=y'p^{-n}q^{-m}r^{-h}y'_0,\\
& v_{p}=\xi(p-q)^m(r-p)^h~\xi_0,\\
& w_{-p}=\eta(q-p)^{-m}(r-p)^{-h}~\eta_0,\\
& s_{a,b}=z' \bigg(\frac{p-a}{p+b}\bigg)^n\bigg(\frac{q-a}{q+b}\bigg)^m
\bigg(\frac{r-a}{r+b}\bigg)^hz'_0+\frac{1}{a+b}, \\
& \sg=\tau(p-q)^{-nm}(q-r)^{-mh}(r-p)^{-nh}\sg_0,
\end{align}
\end{subequations}
where $x_0,~y_0,~y'_0,~\xi_0,~\eta_0,~z'_0,~\sg_0$ are constants.

Under transformations \eqref{dkp-trans}, autonomous discrete KP equations \eqref{dkp-list}
yield deformed equations
\begin{subequations}\label{de-dkp-list}
\begin{align}
\mbox{lpKP equation}:
~~&\wt{x}(\wh{\wt{x}}-\wt{\wb{x}})+\wh{x}(\wh{\wb{x}}-\wh{\wt{x}})+\wb{x}(\wt{\wb{x}}-\wh{\wb{x}})=0,\\
\mbox{lpmKP-I equation}:~~&\frac{\wh{y}-\wt{y}}{\wh{\wt{y}}}+\frac{\wt{y}-\wb{y}}{\wb{\wt{y}}}+\frac{\wb{y}-\wh{y}}{\wh{\wb{y}}}=0, \\
\mbox{lpmKP-II equation}:~~&\frac{1}{\wh{y'}}\frac{(\wt{y'}-\wb{y'}){\wh{\phantom{a}}}}{\wt{y'}-\wb{y'}}
=\frac{1}{\wt{y'}}\frac{(\wh{y'}-\wb{y'}){\wt{\phantom{a}}}}{\wh{y'}-\wb{y'}}
=\frac{1}{\wb{y'}}\frac{(\wt{y'}-\wh{y'}){\wb{\phantom{a}}}}{\wt{y'}-\wh{y'}},\\
\mbox{asymmetric lpmKP-I equation}:~~&\frac{\wt{\xi}}{\wh{\wt{\xi}}}
+\frac{\wt{\xi}}{\wb{\wt{\xi}}}-
\frac{\wh{\xi}+\wb{\xi}}{\wh{\wb{\xi}}}=0, \\
\mbox{asymmetric lpmKP-II equation}:~~&
\frac{\wb{\wh{\eta}}\wt{\eta}}{\wh{\eta}\wb{\eta}}=
\frac{\wh{\wt{\eta}}-\wb{\wt{\eta}}}{\wh{\eta}-\wb{\eta}}, \\
\mbox{lKP-NQC equation}:~~&
\frac{(\wh{z'}-\wh{\wt{z'}})(\wt{z'}-\wt{\wb{z'}})
(\wb{z'}-\wh{\wb{z'}})}{(\wt{z'}-\wh{\wt{z'}})(\wb{z'}-\wt{\wb{z'}})
(\wh{z'}-\wh{\wb{z'}})}=1, \\
\mbox{blKP equation}:~~&
\wh{\wt{\sg}}\wb{\sg}+\wh{\wb{\sg}}\wt{\sg}+\wb{\wt{\sg}}\wh{\sg}=0.
\end{align}
\end{subequations}

\section{List of notations} \label{A:G3}

\begin{itemize}
\item{Diagonal matrices:
\begin{subequations}
\begin{align}
& \Ga^{\tyb{N}}_{\ty{D}}(\{k_i\}^{N}_{1})=\mathrm{Diag}(k_1, k_2, \ldots, k_N), \\
& \Lb^{\tyb{N'}}_{\ty{D}}(\{\ka_j\}^{N'}_{1})=\mathrm{Diag}(\ka_1, \ka_2, \cdots, \ka_{N'}),
\end{align}
\end{subequations}
}
\item{Jordan block matrices:
\begin{subequations}
\begin{align}
& \Ga^{\tyb{N}}_{\ty{J}}(a)
=\left(\begin{array}{cccccc}
a & 0    & 0   & \cdots & 0   & 0 \\
1   & a  & 0   & \cdots & 0   & 0 \\
0   & 1  & a   & \cdots & 0   & 0 \\
\vdots &\vdots &\vdots &\vdots &\vdots &\vdots \\
0   & 0    & 0   & \cdots & 1   & a
\end{array}\right)_{N\times N}, \\
& \Lb^{\tyb{N'}}_{\ty{J}}(b)
=\left(\begin{array}{cccccc}
b & 0    & 0   & \cdots & 0   & 0 \\
1   & b  & 0   & \cdots & 0   & 0 \\
0   & 1  & b   & \cdots & 0   & 0 \\
\vdots &\vdots &\vdots &\vdots &\vdots &\vdots \\
0   & 0    & 0   & \cdots & 1   & b
\end{array}\right)_{N'\times N'},
\end{align}
\end{subequations}
}
\item{Lower triangular Toeplitz matrix \cite{Z-KdV}:
\begin{equation}
\bT^{\tyb{N}}(k_{\mu})
=\left(\begin{array}{cccccc}
\rho_{\mu} & 0    & 0   & \cdots & 0   & 0 \\
\frac{\partial_{k_{\mu}}\rho_{\mu}}{1!} & \rho_{\mu}   & 0   & \cdots & 0   & 0 \\
\frac{\partial^2_{k_{\mu}}\rho_{\mu}}{2!} & \frac{\partial_{k_{\mu}}\rho_{\mu}}{1!}  & \rho_{\mu}  & \cdots & 0   & 0 \\
\vdots &\vdots &\cdots &\vdots &\vdots &\vdots \\
\frac{\partial^{N-1}_{k_{\mu}}\rho_{\mu}}{(N-1)!} & \frac{\partial^{N-2}_{k_{\mu}}\rho_{\mu}}{(N-2)!} &
\frac{\partial^{N-3}_{k_{\mu}}\rho_{\mu}}{(N-3)!}
& \cdots &  \frac{\partial_{k_{\mu}}\rho_{\mu}}{1!}  & \rho_{\mu}
\end{array}\right)_{N\times N},
\label{T}
\end{equation}
}
\item{Skew triangular Toeplitz matrix:
\begin{equation}
\bH^{\tyb{N'}}(\ka_{\nu})
=\left(\begin{array}{ccccc}
\frac{\partial^{N'-1}_{\ka_{\nu}}\vr_{\nu}}{(N'-1)!} & \cdots  & \frac{\partial^{2}_{\ka_{\nu}}\varrho_{\nu}}{2!}  &
\frac{\partial_{\ka_{\nu}}\varrho_{\nu}}{1!}  & \varrho_{\nu}\\
\frac{\partial^{N'-2}_{\ka_{\nu}}\varrho_{\nu}}{(N'-2)!} & \cdots & \frac{\partial_{\ka_{\nu}}\varrho_{\nu}}{1!}  & \varrho_{\nu} & 0\\
\frac{\partial^{N'-3}_{\ka_{\nu}}\varrho_{\nu}}{(N'-3)!} &\cdots & \varrho_{\nu} & 0 & 0\\
\vdots &\vdots & \vdots & \vdots & \vdots\\
\varrho_{\nu} & \cdots & 0 & 0 & 0
\end{array}
\right)_{N'\times N'},
\label{H}
\end{equation}
}
where $\ro_\mu$ and $\vr_\nu$ are discrete exponential functions defined by
\begin{subequations}
\begin{align}
&\ro_\mu=
\prod_{i=n_0}^{n-1}\bigg(1+\frac{k_\mu}{p_i}\bigg)
\prod_{j=m_0}^{m-1}\bigg(1+\frac{k_\mu}{q_j}\bigg)\prod_{l=h_0}^{h-1}\bigg(1+\frac{k_\mu}{r_l}\bigg)\rho^{(0)}_\mu, ~~
\ro^{(0)}_\mu \in \mathbb{C},\\
&\vr_\nu=\prod_{i=n_0}^{n-1}\bigg(1-\frac{\ka_\nu}{p_i}\bigg)^{-1}
\prod_{j=m_0}^{m-1}\bigg(1-\frac{\ka_\nu}{q_j}\bigg)^{-1}\prod_{l=h_0}^{h-1}\bigg(1-\frac{\ka_\nu}{r_l}\bigg)^{-1}\varrho^{(0)}_\nu, ~~
\vr^{(0)}_\nu \in \mathbb{C}.
\end{align}
\end{subequations}
\end{itemize}
Meanwhile, the following expressions need to be considered:
\begin{subequations}\label{notations}
\begin{align}
& N\mbox{\hbox{-}th~vector:}~~\br_{\hbox{\tiny{\it D}}}^{\hbox{\tiny{[{\it N}]}}}(\{k_j\}_{1}^{N})=(\rho_1, \rho_2, \cdots, \rho_N)^{\st},\\
& N'\mbox{\hbox{-}th~vector:}~~\ts_{\hbox{\tiny{\it D}}}^{\hbox{\tiny{[{\it N'}]}}}(\{\ka_j\}_{1}^{N'})=(\varrho_1, \varrho_2, \cdots, \varrho_{N'}),\\
& N\mbox{\hbox{-}th~vector:}~~\br_{\ty{J}}^{\tyb{N}}(k_1)=\Bigl(\rho_1, \frac{\partial_{k_1}\rho_1}{1!},
\cdots, \frac{\partial^{N-1}_{k_1}\rho_1}{(N-1)!}\Bigr)^{\st},\\
& N'\mbox{\hbox{-}th~vector:}~~\ts_{\ty{J}}^{\tyb{N'}}(\ka_1)=\Bigl(\frac{\partial^{N'-1}_{\ka_1}\varrho_1}{(N'-1)!},
\cdots, \frac{\partial_{\ka_1}\varrho_1}{1!},\varrho_1 \Bigr),\\
& N\times N' ~\mathrm{matrix:}~~\bG^{\tyb{N;N'}}_{\ty{DD}}(\{k_i\}^{N}_{1};\{\ka_j\}^{N'}_{1})
=(g_{i,j})_{N\times N'},~~~g_{i,j}=\frac{1}{k_i+\ka_j},\\
& N_1\times N'_2 ~\mathrm{matrix:}~~\bG^{\tyb{N$_1$;N$_2$'}}_{\ty{DJ}}(\{k_i\}^{N_1}_{1};d)
=(g_{i,j})_{N_1\times N'_2},~~~g_{i,j}=-\Bigl(\frac{-1}{k_i+d}\Bigr)^j,\\
& N_2\times N'_1 ~\mathrm{matrix:}~~\bG^{\tyb{N$_2$;N$_1$'}}_{\ty{JD}}(c;\{\ka_j\}^{N'_1}_{1})
=(g_{i,j})_{N_2\times N'_1},~~~g_{i,j}=-\Bigl(\frac{-1}{c+\ka_j}\Bigr)^i,\\
& N_1\times N'_2 ~\mathrm{matrix:}~~\bG^{\tyb{N$_1$;N$_2$'}}_{\ty{JJ}}(c;d)
=(g_{i,j})_{N_1\times N'_2},~~~g_{i,j}=\mathrm{C}^{i-1}_{i+j-2}\frac{(-1)^{i+j}}{(c+d)^{i+j-1}},
\end{align}
\end{subequations}
where \[\mathrm{C}^{i}_{j}=\frac{j!}{i!(j-i)!},~~(j\geq i).\]

\end{appendix}


{\small
\begin{thebibliography}{99}


\bibitem{H-BLKP} R. Hirota,
        Discrete analogue of a generalized Toda equation,
        J. Phys. Soc. Jpn., 50(11), 1981, 3785-3791.
\bibitem{Miwa} T. Miwa,
        On Hirota's difference equations,
        Proc. Japan Acad. Ser. A Math. Sci., 58(1), 1982, 9-12.
\bibitem{D-1982} E. Date, M. Jimbo, T. Miwa,
        Method for generating discrete soliton equations. V,
        J. Phys. Soc. Jpn., 52, 1983, 766-771.
\bibitem{N-KP-DL} F.W. Nijhoff, H.W. Capel, G.L. Wiersma, G.R.W. Quispel,
        B\"{a}cklund transformations and three-dimensional lattice equations,
        Phys. Lett. A, 105(6), 1984, 267-272.
\bibitem{N-KP} F.W. Nijhoff,
        {\it Discrete Systems and Integrability},
        Math5492, University of Leeds, 2010.
\bibitem{WZ} W. Feng, S.L. Zhao,
        Generalized Cauchy matrix approach for lattice KP-type equations,
        Commun. Nonlinear Sci. Numer. Simulat., 18(7), 2013, 1652-1664.
\bibitem{ABS-2012} V.E. Adler, A.I. Bobenko, Yu.B. Suris,
        Classification of integrable discrete equations of octahedron type,
        Int. Math. Res. Notices, 2012(86), 2012, 1822-1889.
\bibitem{SC} R. Sahadevana, H.W. Capel,
        Complete integrability and singularity confinement of nonautonomous modified Korteweg-de Vries and sine Gordon mappings,
        Physica A, 330, 2003, 373-390.
\bibitem{SRH} R. Sahadevan, O.G. Rasin, P.E. Hydon,
        Integrability conditions for nonautonomous quad-graph equations,
        J. Math. Anal. Appl., 331, 2007, 712-726.
\bibitem{GR} B. Grammaticos, A. Ramani,
        Singularity confinement property for the (non-autonomous) Adler-Bobenko-Suris integrable lattice equations,
        Lett. Math. Phys., 92, 2010, 33-45.
\bibitem{WTS} R. Willox, T. Tokihiro, J. Satsuma,
        Darboux and binary Darboux transformations for the nonautonomous discrete KP equation,
        J. Math. Phys., 38, 1997, 6455-6469.
\bibitem{WTS-1} R. Willox, T. Tokihiro, J. Satsuma,
        Nonautonomous discrete integrable systems,
        Chaos, Solitons Fractals, 11, 2000, 121-135.
\bibitem{H-cr} M. Hay,
        Casorati determinant solutions to the non-autonomous cross-ratio equation,
        http://lsec.cc.ac.cn/~icnwta2/postersession/hay.pdf, 2010.
\bibitem{KM} K. Kajiwara, A. Mukaihira,
        Soliton solutions for the non-autonomous discrete-time Toda lattice equation,
        J. Phys. A: Math. Gen., 38, 2005, 6363-6370.
\bibitem{KO-1} K. Kajiwara, Y. Ohta,
        Bilinearization and Casorati determinant solution to the non-autonomous discrete KdV equation,
        J. Phys. Soc. Jpn., 77, 2008, 054004 (9pp).
\bibitem{KO-2} K. Kajiwara, Y. Ohta,
        Bilinearization and Casorati determinant solutions to non-autonomous 1+1 dimensional discrete soliton equations,
        RIMS K\^{o}ky\^{u}roku Bessatsu, B13, 2009, 53-73.
\bibitem{SZZ} Y. Shi, D.J. Zhang, S.L. Zhao,
        Solutions to the non-autonomous ABS lattice equations: Casoratians and bilinearization (in Chinese),
        Sci. Sin. Math., 44(1), 2014, 37-54, arXiv:1201.6478.
\bibitem{NZ} L.J. Nong, D.J. Zhang,
        Non-autonomous discrete Boussinesq equation: Solutions and consistency,
        Chin. Phys. B, 23(7), 2014, 070202 (6pp).
\bibitem{S-1884} J. Sylvester,
        Sur l'equation en matrices $px=xq$,
        C. R. Acad. Sci. Paris, 99, 1884, 67-71, 115-116.
\bibitem{BR-BLMS-1997} R. Bhatia, P. Rosenthal,
        How and why to solve the operator equation $AX-XB=Y$,
        Bull. London Math. Soc., 29, 1997, 1-21.
\bibitem{ZZ} D.J. Zhang, S.L. Zhao,
        Solutions to ABS lattice equations via generalized Cauchy matrix approach,
        Stud. Appl. Math., 131, 2013, 72-103.
\bibitem{ZSF} S.L. Zhao, S.F. Shen, W. Feng,
        Solutions to Kadomtsev-Petviashvili system: generalized Cauchy matrix approach,
        arXiv:1404.3043v2, 2014.
\bibitem{N-KP-CL} G.L. Wiersma, H.W. Capel,
        Lattice equations, hierarchies and Hamiltonian structures,
        Physica A, 149, 1988, 49-74.
\bibitem{Z-KdV} D.J. Zhang, Notes on solutions in Wronskian form to soliton equations: KdV-type,
        arXiv:nlin.SI/0603008, 2006.

\end{thebibliography}}
\end{document}